\documentclass[fleqn,usenatbib]{mnras}

\usepackage{newtxtext,newtxmath}
\usepackage{booktabs}
\usepackage{threeparttable}

\usepackage[T1]{fontenc}

\DeclareRobustCommand{\VAN}[3]{#2}
\let\VANthebibliography\thebibliography
\def\thebibliography{\DeclareRobustCommand{\VAN}[3]{##3}\VANthebibliography}

\usepackage{graphicx}	
\usepackage{amsmath}	
\usepackage{booktabs}
\usepackage{xcolor}
\usepackage{ulem}

\def\sec#1{Section~\ref{sec:#1}}
\def\app#1{Appendix~\ref{sec:#1}}
\def\Fig#1{Figure~\ref{fig:#1}}
\def\fig#1{Figure~\ref{fig:#1}}
\def\Table#1{Table~\ref{tab:#1}}
\def\tab#1{Table~\ref{tab:#1}}
\def\eq#1{eq.~(\ref{eq:#1})}
\def\Eq#1{Eq.~(\ref{eq:#1})}

\newcommand{\ap}{{\scriptsize AUTOPROF }}
\newcommand{\pym}{{\scriptsize PYMORPH }}

\newcommand{\kms}{\ifmmode\,{\rm km}\,{\rm s}^{-1}\else km$\,$s$^{-1}$\fi}
\newcommand{\Rd}{\ifmmode\,R_{\rm d}\else $R_{\rm d}$\fi}
\newcommand{\be}{\begin{equation}}
\newcommand{\ee}{\end{equation}}
\newcommand\ltsima{$\; \buildrel < \over \sim \;$}
\newcommand\ltsim{\lower.5ex\hbox{\ltsima}}
\newcommand\gtsima{$\; \buildrel > \over \sim \;$}
\newcommand\gtsim{\lower.5ex\hbox{\gtsima}}

\newcommand{\magss}{\ifmmode {{{{\rm mag}~{\rm arcsec}}^{-2}}}
             \else {{{mag}$~${arcsec}$^{-2}$}}
             \fi}

\def \ion#1#2{#1{\footnotesize{#2}}\relax}
\def \hi{\ion{H}{I}}
\def \littleprime{\ifmmode{\scriptscriptstyle \prime }
     \else{\hbox{$\scriptscriptstyle \prime$ }}\fi}
\def \arcsec{\raise .9ex \hbox{\littleprime\hskip-3pt\littleprime}}

\title[MaNGA galaxy properties]{MaNGA galaxy properties -- I. An extensive optical, mid-infrared photometric, and environmental catalogue}

\author[Arora et al.]{
Nikhil Arora,$^{1}$\thanks{E-mail: nikhil.arora@queensu.ca}
Connor Stone,$^{1}$
Stéphane Courteau$^{1}$
and Thomas H. Jarrett$^{2}$
\\
$^{1}$Department of Physics, Engineering Physics and Astronomy, Queen's University, Kingston, ON K7L 3N6, Canada\\
$^{2}$Astronomy Department, University of Cape Town, Rondebosch 7701, South Africa\\
}

\date{Accepted XXX. Received YYY; in original form ZZZ}

\pubyear{2021}

\begin{document}
\defcitealias{RC15}{RC15}
\defcitealias{Z17}{Z17}
\defcitealias{Fischer2019}{F19}
\defcitealias{Pace2019}{P19}
\defcitealias{Benito2019}{B19}
\defcitealias{Gilhuly2018}{GC18}

\label{firstpage}
\pagerange{\pageref{firstpage}--\pageref{lastpage}}
\maketitle

\begin{abstract}
    We present an extensive catalogueue of non-parametric structural properties derived from optical and mid-infrared imaging for 4585 galaxies from the MaNGA survey.
    DESI and Wide-field Infrared Survey (WISE) imaging are used to extract surface brightness profiles in the $g, r, z, {\it W1}, W2$ photometric bands.
    Our optical photometry takes advantage of the automated algorithm \ap and probes surface brightnesses that typically reach below 29 $\magss$ in the $r$ band, while our WISE photometry achieves 28 $\magss$ in the ${\it W1}$ band.
    Neighbour density measures and central/satellite classifications are also provided for a large subsample of the MaNGA galaxies.
    Highlights of our analysis of galaxy light profiles include
    (i) an extensive comparison of galaxian structural properties that illustrates the robustness of non-parametric extraction of light profiles over parametric methods;
    (ii) the ubiquity of bimodal structural properties, suggesting the existence of galaxy families in multiple dimensions; and 
    (iii) an appreciation that structural properties measured relative to total light, regardless of the fractional level, are uncertain.
    We study galaxy scaling relations based on photometric parameters, and present detailed comparisons with literature and theory.
    Salient features of this analysis include the near-constancy of the slope and scatter of the size--luminosity and size--stellar mass relations for late-type galaxies with wavelength, and the saturation of the central surface density, measured within 1\,kpc, for elliptical galaxies with $M_* > 10.7\,M_{\odot}$ 
    (corresponding to $\Sigma_1 \simeq 10^{10}\,M_{\odot}\,{\rm kpc}^{-2}$).
    The multiband photometry, environmental parameters, and structural scaling relations presented are useful constraints for stellar population and galaxy formation models.  
\end{abstract}



\section{Introduction} \label{sec:intro}

A comprehensive picture of galaxy formation and evolution, as well as reliable data-model comparisons, requires access to large homogeneous photometric and spectroscopic samples of galaxies covering a broad range of morphologies, stellar and dynamical masses, star formation histories, environmental conditions, and more.
Previously, such measurements have been extracted from large imaging and spectroscopic wide-field surveys such as 2MASS \citep{Jarrett2000}, SDSS \citep{York2000}, GAMA \citep{gama}, and others. 
More recently, large-scale acquisition of spatially resolved structural properties of galaxies has been achieved with integral-field spectroscopy. 
The latter is especially valuable in this context since it allows the synchronous collection of global and spatially resolved chemical and dynamical properties such as stellar mass surface density, age, metallicities, star formation rates, circular velocity, velocity dispersion and more, across a broad wavelength range. 
Pioneering integral-field spectroscopy surveys of galaxies such as SAURON \citep{Bacon2001}, ATLAS$\rm ^{3D}$ \citep{atlas3d}, CALIFA \citep{Walcher2014}, and SAMI \citep{sami}, provide such properties, although for relatively small samples.

With its spatially resolved optical spectroscopic data for ${\sim}10000$ nearby galaxies ($ z{\sim}0.15$), the SDSS-IV survey ``Mapping Nearby Galaxy at Apache point observatory" (MaNGA; \citealp{Bundy2015, Wake2017}) has heralded a new era of large-scale galaxy studies with integral field units (IFUs)\footnote{While the MaNGA target sample includes $\sim$10,000 objects, 4585 of them are publicly available at the time of writing.}.
MaNGA uses a 127-fiber IFU to provide high-quality spatially resolved kinematic and chemical information for galaxies by collecting spectra at each pixel out to a maximum of 1.5/2.5 effective radii ($R_{\rm eff}$) for 85/15 percent of the MaNGA sample~\citep{Cano2016, Sanchez2016, Sanchez2018, Graham2018, Mendez2018}.
The MaNGA galaxies were selected from the NSA \footnote{NASA Sloan Atlas, \cite{Blanton2011}} catalogue with no inclination or size selection cuts.
For more details about the MaNGA sample, see \cite{Bundy2015} and \cite{Wake2017}.

Large-scale IFU surveys, such as MaNGA and CALIFA \citep{Walcher2014}, open an avenue for the exploration of spatially resolved counterparts to global structural drivers and scaling relations in galaxies.
For instance, the MaNGA survey has already yielded new insights about evolutionary mechanisms such as star formation quenching; finding it to be regulated on global galaxy scales while star formation density is controlled locally \citep{Bluck2020}. 
Using the Pipe3D \citep{Sanchez2018} output for MaNGA, \cite{Menguiano2020} have also explored the relation between the average stellar age and gas-phase metallicity for star-forming systems.
While this relation had been observed globally \citep{Lian2015}, the spatially resolved nature of the MaNGA data also revealed a local correlation.

In addition to the published spatially resolved spectroscopic MaNGA data, a comprehensive understanding of MaNGA galaxies requires a full suite of structural and environmental galaxy properties.
A catalogue of model-dependent MaNGA galaxy structural parameters (effective sizes, total magnitudes, S\'ersic indices, ellipticities, position angles) already exists \citep[][hereafter \citetalias{Fischer2019}]{Fischer2019}.
The MaNGA \pym Photometric Value Added Catalogue (MPP-VAC) provides structural parameters for MaNGA galaxies extracted through S\'ersic and S\'ersic+Exponential fits to 2D surface brightness profiles.
However, model-dependent assumptions about the light distribution of galaxies severely bias and limit any investigations of galaxy structure and evolution, as shown below. 
Providing galaxy structural parameters in {\it non-parametric} fashion is the main goal of this study.

In this paper, we present an extensive compilation of {\it model-independent} galaxian structural parameters based on optical and mid-infrared (MIR) imaging of all MaNGA galaxies extracted respectively from the Dark Energy Sky Instrument Legacy Imaging Survey \citep{desi, Dey2019}\footnote{In what follows, the acronym DESI is meant to represent the Dark Energy Sky Instrument Legacy Imaging Survey.} and Wide-field Infrared Survey Explorer ({\it WISE}) surveys.
The combination of optical and MIR bands is especially sensitive to the evolved stellar population, sites of star formation, and distribution of dust within galaxies.
Broader multiband coverage for MaNGA galaxies is also being developed elsewhere; 
major imaging campaigns of MaNGA galaxies at radio \citep{manga_h1} and ultraviolet wavelengths (\citealp{manga_uv}) are either in place or in progress.
These will expand our understanding of MaNGA galaxies with clearer views of processes such as star formation on small spatial/time scales and the distribution of neutral hydrogen within galaxies, the fuel for star formation.

The surface brightness (surface brightness) profiles extracted from the DESI and WISE imaging surveys for the MaNGA galaxies allow us to extract numerous non-parametric structural properties representing various measures of galaxy size, surface brightness, luminosity, stellar mass, and stellar mass surface density.
Along with the structural parameters for MaNGA galaxies, environmental demographics are also provided below. 
Our compilation supplements fifth nearest-neighbour density measurements for MaNGA galaxies \citep[see][for details]{GEMA2015, Etherington2015}, with important information about the gravitational dominance of a galaxy (i.e. central and satellite designation) within its own halo.
These considerations motivate the assembly of a catalogue of environmental properties for a large subsample of MaNGA galaxies, in addition to the photometric catalogues. 
The environmental catalogue that we present below includes measures of neighbour densities and classification of these galaxies as a central or satellite.

A natural outcome of such an extensive data compilation is the construction and study of galaxy scaling relations. 
The slope, zero-point, and scatter of such empirical scaling relations present evidence of the underlying physics that dictates structure formation and evolution of galaxies \citep{Mo1998, Courteau2007, Dutton2007, Brook2012, Hall2012, Lange2015}.
The MaNGA non-parametric data sets produced in this study allow for an analysis of various galaxy scaling relations in different photometric bands. 
Among others, the size--stellar mass ($R-M_*$) relation and the stellar surface density ($\Sigma_1$ measured within 1 kpc) - stellar mass relations ($\Sigma_1-M_*$) 
are presented. 
As discussed below, these relations enable constraints of the gas accretion and merger history as well as quenching and feedback models in galaxies \citep{Chiosi2002,Shen2003,Franx2008,Saglia2010,Fang2013,Woo2019}.

This paper is organized as follows: in \sec{data}, we describe the procedures used to extract surface brightness profiles in optical and MIR bands.
The extraction and correction of non-parametric galaxy structural properties are also discussed and presented in tabular format.
The environmental catalogue is presented in \sec{environment}. 
With our catalogue in place, we provide in \sec{param_dist} a broad overview of the DESI and WISE parameters for our MaNGA Public sample.  
Unimodal and bimodal structural distributions are also highlighted. 
In \sec{compare}, we compare our model-independent structural parameters for MaNGA galaxies, such as size, apparent magnitude, and stellar mass, with literature values. 
Our MIR photometry and derived structural properties are compared to optical structural properties in \sec{ir_photo}.
With our optical and MIR photometry validated, we combine these data sets in \sec{sr} to infer structural galaxy scaling relations of MaNGA galaxies with a focus on the size--stellar mass, $R-M_*$, relation in \sec{sizemass}, and the central surface density--stellar mass, $\Sigma_1 - M_*$, relation in \sec{sigmamass}. 
We conclude in \sec{conclusion}.

\section{Light Profile Extraction} \label{sec:data}

Our galaxy sample consists of 4585 galaxies from the public release of the MaNGA survey. 
The full MaNGA galaxy survey will yield $\sim$10\,000 objects with a uniform distribution of stellar mass $M_* > 10^9 {\rm M_{\odot}}$, with no inclination or size selections.

For our optical photometry, we have cross-correlated the public MaNGA data release with the DESI \footnote{\url{https://www.legacysurvey.org}} \citep{desi} and extracted $\rm 10\times 10\,{arcmin^2}$ images in the \textit{g, r,} and \textit{z} bands for 4585 matching galaxies.
The large galaxy images ensure that the background sky level can be robustly characterized and subtracted.
We use our fully automated `\ap' code (\sec{autoprof}) to extract azimuthally averaged surface brightness profiles from DESI {\it g-,r-} and {\it z-} band images, with the solution fit to the {\it r} band for its minimal dust extinction and high signal-to-noise ratio (S/N; and for consistency with the parallel Photometry and Rotation Curve Observations from Extragalactic Surveys, or ``PROBES'', investigation by \citealt{Stone2019} and \citealt{Stone2020}). 
The $r$-band isophotal solution is applied to the $g$ and $z$ images for uniformity of the position angles, ellipticities, fluxes and color gradients. 
Unless otherwise stated, all band-dependent structural parameters refer to the $z$ band photometry.  
The automation of \ap is well suited to large surveys like DESI, where interactive surface brightness extraction methods, such as those based on the {\scriptsize {\scriptsize XVISTA}} data reduction package\footnote{\url{http://ganymede.nmsu.edu/holtz/xvista/}}, become prohibitive~\citep[see][for more details]{Courteau1996,McDonald2011,Hall2012,Gilhuly2018}.

Our MIR surface brightness profiles are derived from the WISE Large Galaxy Atlas and the Extended Source catalogue (WXSC; \citealt{Jarrett2019})\footnote{In what follows, WISE and the WISE Large Galaxy Atlas and the Extended Source catalogue are taken as synonymous \citep{Jarrett2019}.}, which features custom image mosaic construction to produce native angular resolution products that include both WISE and NEOWISE imaging \footnote{(NEO)WISE data imaging and catalogueues can be retrieved from \url{https://irsa.ipac.caltech.edu/}}to improve the sensitivity; 
complete details of image construction are provided in \cite{Jarrett2012}.
Here we use the {\it W1} (3.4$\rm \mu$m) and {\it W2} (4.6$\rm \mu$m) mosaics with 5.9$\arcsec$ and 6.5$\arcsec$ spatial resolution, respectively. 
These bands are sensitive to the evolved stellar populations \citep{Jarrett2013}, and hence the stellar mass content and distribution for the target galaxies.
We have extracted from the WISE surface brightness profiles the same structural properties as those measured from the DESI optical photometry, except for the stellar surface density within 1 kpc which is not resolved for many WISE profiles.
The robustness of our WISE derived structural properties is demonstrated in \sec{ir_photo}.
Among others, the {\it W1} and {\it W2} band passes are especially sensitive to older, more mass dominant stellar populations resulting in robust stellar mass measurements.
The comparison of stellar masses from our DESI and WISE photometry is presented in \sec{ir_photo}.
In what follows, all surface brightnesses used and reported are in the AB magnitude system.

In \app{profSB}, we provide \Table{desi_profile} and \ref{tab:wise_profile} to present the output format of the public multiband surface brightness profiles.

\subsection{\ap}\label{sec:autoprof}

\begin{figure*}
    \centering
    \includegraphics[width=\linewidth]{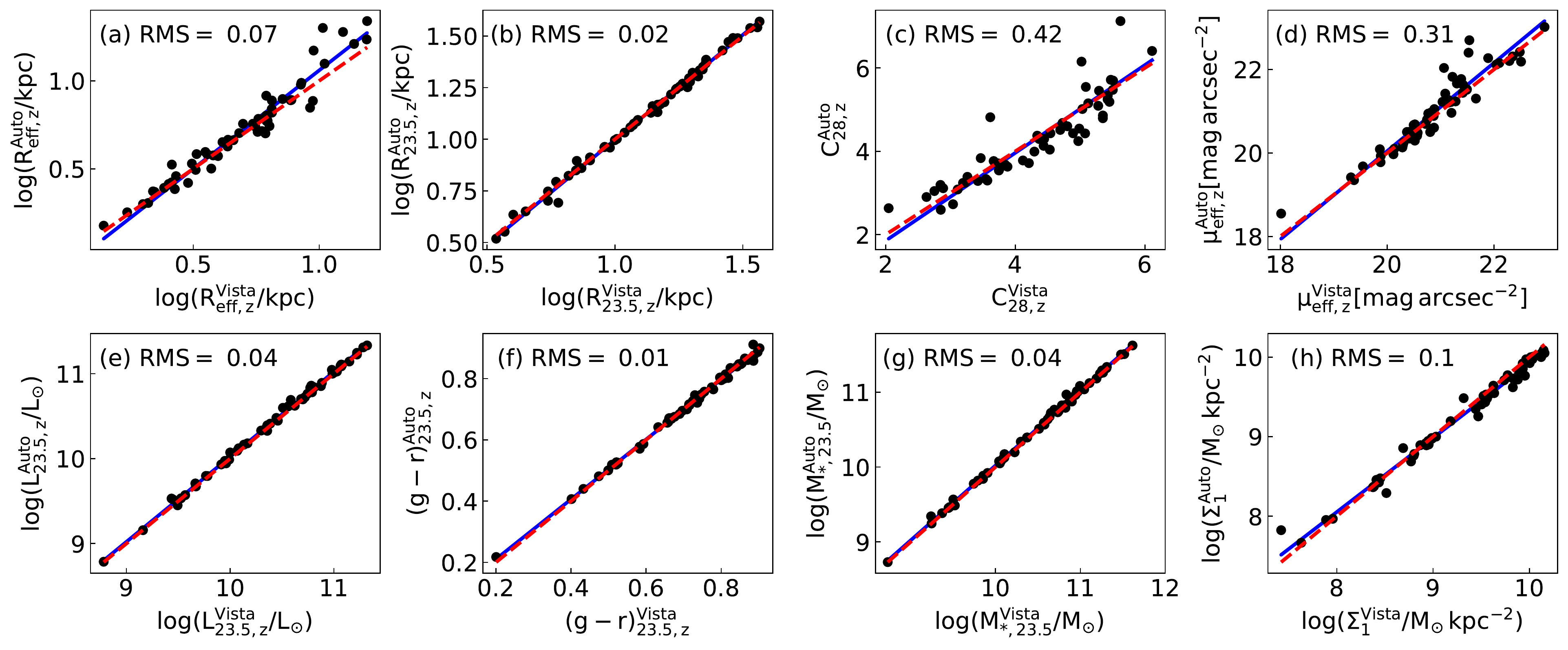}
    \caption{Comparison of photometrically derived non-parametric quantities from {\scriptsize XVISTA} interactive routines and \ap with: (panel a) effective radius [$R_{\rm eff}$], (panel b) isophotal radius [$R_{\rm 23.5}$], (panel c) concentration index [$C_{\rm 28}$], (panel d) effective surface brightness [$\mu_{eff}$], (panel e) luminosity within $R_{\rm 23.5}$ [$L_{\rm 23.5}$], (panel f) colour within $R_{\rm 23.5}$ [$(g-r)_{\rm 23.5}$], (panel g) stellar mass within $R_{\rm 23.5}$ [$M_*$], and (panel h) stellar mass surface density within 1 kpc [$\Sigma_1$]. The blue solid line shows the best orthogonal linear fit and the red dashed line shows the one-to-one line. All measurements are from DESI \textit{z}-band photometry. The text in the top left-hand panel presents root mean square (rms) of the residuals for the one-to-one line.}
    \label{fig:vista_auto_compare}
\end{figure*}

The automated surface brightness profile extraction algorithm, ``\ap'' works on a combination of standard and machine learning based numerical techniques.
A complete description of the \ap pipeline is presented in Stone et al (in preparation).
What follows is a brief outline.

The \ap pipeline first computes basic image parameters such as the background sky level, PSF, and galaxy centre.
The background is determined as the mode of the pixel flux distribution.
The PSF is determined uniquely for each image with circular apertures placed on a set of 50 non-saturated stars found using an edge detection convolutional filter.
The galaxy centre is found by starting at the centre of the image and iteratively moving to brighter regions until a peak is found.
The centre finding iterative step works by following the direction of increasing brightness as determined by the first FFT coefficient phase~\citep[FFT:][]{cooley1965} of the flux values around a circular aperture.
On the sky subtracted and centred image, a global position angle and ellipticity is fit to the galaxy at the outer region of the galaxy (at approximately 3$\sigma$ above the sky noise).
The global fit is performed by minimizing the power in the second FFT coefficient of the flux values along the isophote.
An elliptical isophote solution is fit to the  sky-subtracted and centred image by simultaneously minimizing the second Fourier coefficient and a regularization term~\citep{shalev2014}.
The regularization term is the $l_1$ norm of the difference in ellipticity and position angle between adjacent isophotes.
The use of a regularization term is borrowed from machine learning; other automated surface brightness profile extraction techniques take advantage of machine learning for all steps~\citep{Tuccillo2018, Smith2020}.

A surface brightness profile is extracted at incremental radii corresponding to the median flux along each isophote. 
For isophotes at larger radii, \ap extracts a non-overlapping ``band'' of pixel flux values from the image, thus yielding a higher S/N for each isophote.
A curve of growth is then calculated by integrating the surface brightness profile appropriately. 
The surface brightness uncertainty is computed from the 68.3\% quartile range of the sampled flux along each isophote; this uncertainty is also propagated to the curve of growth.

Other surface brightness profile extraction techniques exist, such as the interactive package PROFILE in {\scriptsize XVISTA} \citep{Courteau1996,McDonald2011,Hall2012,Gilhuly2018}.
\Fig{vista_auto_compare} shows a comparison of several (uncorrected) structural properties defined in \sec{parameter} for $\sim$50 MaNGA galaxies extracted using {\scriptsize XVISTA} and \ap.
A good match between the two methods is found for $R_{\rm 23.5}$, $L_{23.5}$, $g-r$, and $M_*$.
Other quantities in \Fig{vista_auto_compare}, such as $R_{\rm eff}$, $C_{28}$ (see equation \ref{eq:c28}), and $\mu_{\rm eff}$ (measured at $R_{\rm eff}$), do not match as well. 
The effective radius comparison has a scatter of 0.07 dex, which matches the $R_{\rm eff}$ comparison of SDSS imaging parameters in \citet[][hereafter \citetalias{Gilhuly2018}]{Gilhuly2018} (0.06 dex) with the photometry of \citet{Walcher2014} for CALIFA galaxies. 
An equally poor match for $R_{\rm eff}$ relative to SDSS Petrosian radii is shown in fig. 9 of \citet{Hall2012}.
We also find a large rms of 0.42\,dex comparing $C_{28}$ estimates which reflects the large uncertainties involved in measuring outer radii based on total light \citep{Graham2001, Trujillo2001}.
Likewise, a large rms of 0.42\,dex is found for comparisons of $C_{28}$ estimates, reflecting the large uncertainties involved in measuring outer radii based on total light \citep{Graham2001, Trujillo2001}.
We argue in \sec{compare} that the poor reproducibility of $\mu_{\rm eff}$, $R_{\rm eff}$, and $C_{\rm 28}$, results largely from the ambiguous definition of total integrated luminosity in a photometric band, which these quantities all depend on.
Still, we include these quantities in our tables for comparison with the literature.

\begin{figure*}
    \centering
    \includegraphics[width=\linewidth]{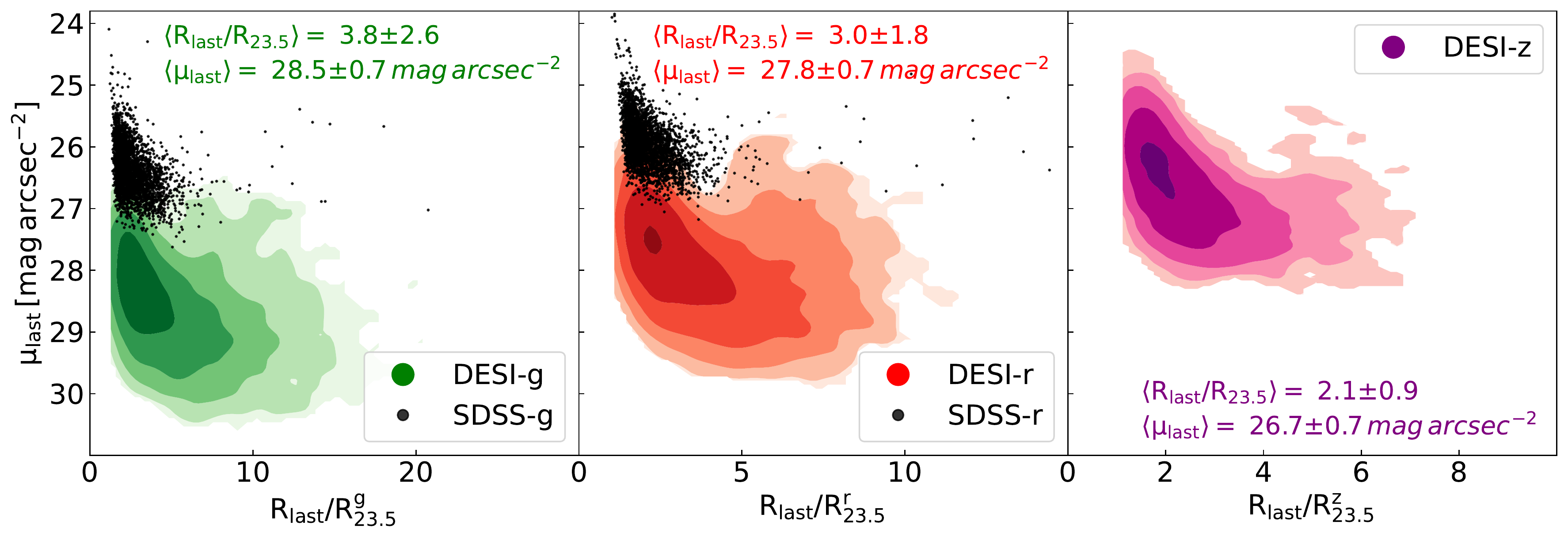}
    \caption{Lowest surface brightness levels and maximal radial extents of DESI surface brightness profiles in the \textit{g} (left-hand panel), \textit{r} (center panel) and \textit{z}-bands (right-hand panel) with coloured contours in log density versus similar quantities for SDSS surface brightness profiles (black dots in the $g$ and $r$ bands; the $z$ band is unavailable) for the same MaNGA galaxies.  The radial axis shows the extent of the last point in each surface brightness profile scaled in units of R$\rm _{23.5}$. The text within each panel indicates the average depth of our DESI photometry in the radius and surface brightness. }
    \label{fig:photo_depth}
\end{figure*}

\Fig{photo_depth} offers a visualization of the greater depth of the DESI imaging (coloured contours) relative to SDSS (black dots for the $g$ and $r$ bands).
The photometric depth is chosen as the point at which a profile reaches an surface brightness error threshold of 0.22~\magss.
As a result, the surface brightness levels are typically $\sim$2~\magss deeper where $\sim$1.5~\magss is attributed to the deeper DESI imaging (over SDSS) and $\sim$0.5~\magss is attributed to the isophote sampling method in \ap (over {\scriptsize XVISTA})\footnote{The error threshold adopted by \citet{Gilhuly2018} for SDSS surface brightness profiles of CALIFA galaxies at $g$, $r$, and $i$ bands is $0.15~\magss$. A typographical error in that paper incorrectly lists that threshold as $1.5~\magss$ (sic) in their Section 3.}

The coloured contours show the surface brightness levels and radial extents at which the surface brightness truncation occurs in the three DESI-{\it grz} bands. 
On average, the DESI photometric solutions extend radially beyond $2(z)-3.8(g)~R_{\rm 23.5}$ (greater radial extent at bluer wavelengths) and can reach surface brightness levels as low as 26.7({\it z}) to 28.5({\it g})~$\magss$, with some profiles going down to 30\,\magss.
Comparison with the SDSS data (extracted by ourselves using {\scriptsize XVISTA} procedures) shows that the combination of DESI images and the \ap surface brightness extraction yields profiles that reach typically ${\sim}2~\magss$ deeper than SDSS analysed with {\scriptsize XVISTA} given the same surface brightness uncertainty.

\subsection{Parameter Extraction} \label{sec:parameter}

We briefly describe the parameters inferred from optical and MIR imaging presented in the photometric and environment catalogues.
A complete listing of our extracted parameters and their units is given in \Table{phototable} and \ref{tab:manga_env}.

\subsubsection{Sizes}\label{sec:size}
Some galaxy sizes are calculated relative to total light or measured at different isophotal levels. 
Relative sizes are calculated at 20, 50, and 80 per cent of the total light.
The errors on these sizes are calculated by standard error propagation from the curve of growth (COG).
Isophotal sizes are measured at the $\rm 22.5$, $\rm 23.5$ and $\rm 25\, \magss$ isophotes from the surface brightness profiles in their respective band. 
Errors on isophotal sizes are calculated by standard error propagation from the surface brightness profile.
While surface brightness errors tend to be larger than COG errors, the relative slope of the profiles results in larger errors for the sizes based on the COG.

The projected sizes $(R)$ extracted from our photometry can be transformed into 3D physical radii ($r$), i.e., $r(R)$, by modelling projection effects.  
For flat discy stellar systems, the physical and projected radii are  interchangeable (e.g. $r(R)\approx R$).
For trixial systems, models suggest that $r(R)\approx (4/3)R$ \citep{Hernquist1990, Ciotti1991,Ouellette2017}.
In \sec{sr}, we review the impact of this transformation on galaxy scaling relations such as the size--mass relation.

\subsubsection{Brightness, luminosities, and colours}
\label{sec:light}

Total apparent magnitudes and total luminosities enclosed within the $R_{\rm 23.5}$ and $R_{\rm 25}$ radii are presented.
The colour terms $g-r$, $g-z$ and $r-z$ are evaluated at $R_{\rm 23.5}$ and $R_{\rm 25}$ in the {\it z} band.

The ellipticity of the last measured isophote of a galaxy is used as the representative ellipticity of that galaxy on the sky.

These quantities are all calculated using linear interpolations of consecutive surface brightness profile or curve of growth data.
If an interpolation is not possible, the last 25 percent of the profile is used to extrapolate up to that point using a linear fit.
A flag for each structural parameter indicates an interpolation or extrapolation.

\subsubsection{Concentration}
The concentration of light, $C_{\rm 28}$, is a ratio of radii that enclose 20 per cent and 80 per cent of the total light: 
\be
    C_{28} = 5\log(R_{80}/R_{20}).
    \label{eq:c28}
\ee

A second measure of concentration, $C_{\rm 25}$, is also presented.  
Its use of $R_{\rm 20}$ and $R_{\rm 50}$ is operationally similar to \Eq{c28}.  
$\rm R_{\rm 50}$ is the effective radius, equivalent to $R_{\rm eff}$. 

\subsubsection{Stellar mass and surface stellar density}\label{sec:stellarmass}

Stellar masses can be inferred from observed colours using stellar mass-to-light colour relations (MLCRs;\citealt{Courteau2014}). 
We use five different stellar mass estimates from the MLCRs presented in \citet[][hereafter \citetalias{RC15}]{RC15}, \citet[][hereafter \citetalias{Z17}]{Z17}, and \citet[][hereafter \citetalias{Benito2019}]{Benito2019}.
These MLCRs differ in their choice of stellar population synthesis model, adopted galactic extinctions, and input data used to calibrate the MLCR (global vs. resolved SED fits ).
Our stellar mass estimates use $g-r$ and $g-z$ colours measured at R$_{23.5}$ in the DESI-{\it z} band.
The use of multiple colours yields better constraints on mass-to-light ratios (\citetalias{RC15}, \citetalias{Z17}, \citealt{Gilhuly2018}).
The mass-to-light ratios ($\rm \Upsilon_*$) calculated from these colours are multiplied by the luminosity in the {\it g, r,} and {\it z} bands measured at R$_{23.5}$, which itself is inferred in the DESI-{\it z} band.
This results in 30 stellar mass estimates that are averaged to provide a stellar mass estimate used throughout this study.
The error in the stellar mass is the standard deviation of the 30 stellar mass measurements.
MLCRs in \citetalias{RC15} and \citetalias{Z17} have only been calibrated for late-type systems and are exclusively applied to those systems. 
The stellar mass measurements for ETGs are thus measured using the MLCR presented in \citetalias{Benito2019}.

The stellar $\rm \Upsilon_*$ enable the direct conversion of COGs into stellar surface density profiles. 
We calculate the stellar surface density within 1\,kpc ($\Sigma_{1}= M_{\rm *,1kpc}/\pi$) and the effective radius ($\Sigma_{\rm eff} = M_{*}/\pi R_{\rm eff}^2$) where $M_*$ is the total stellar mass of the galaxy and  $\rm M_{\rm *,1kpc}$ is the stellar mass within the 1\,kpc aperture.

The MIR stellar masses derived from ${\it W1}-W2$ colours and the MLCR of \cite{Cluver2014} are shown as follows:
\be
    \log(M_*/L_{{\it W1}})= -0.17-2.54\times ({\it W1}-W2), 
    \label{eq:ml_ir}
\ee
where $L_{{\it W1}}=10^{-0.4(M_{{\it W1}}-M_{\odot, {\it W1}})}$ is the luminosity, $M_{{\it W1}}$ is the absolute magnitude in the ${\it W1}$ filter, and $M_{\odot, {\it W1}}=3.24$ is the absolute magnitude of the Sun in ${\it W1}$ band.
The same expression is used to get the luminosity for the WISE $W2$ photometric band, except that $M_{\odot, W2}=3.27$.

These MLCRs use a Chabrier IMF, star formation and AGN activity, dust content, old stellar content, along with detailed stellar mass calibrations \citep{Taylor2011}.
The colour in \eq{ml_ir} is limited to the range $-0.05 \leq ({\it W1} - W2) \leq 0.2$ as redder (bluer) sources may be contaminated by AGN or starburst activity \citep{Jarrett2013}.

\subsubsection{Environmental parameters}\label{sec:environment}

Various environmental properties for 3207 MaNGA galaxies are presented in \Table{manga_env}.
The number of neighbours is extracted from the modified catalogue of \cite{wilman2010} based on SDSS-DR7 \citep{sdssdr7}.
Galaxies are counted as neighbours if they fall within projected circular apertures of varying radii (0.25-3.0 Mpc) and their heliocentric Hubble flow velocities are within $\pm$1500~\kms\ of each other. 

Central and satellite galaxies are identified via a stellar mass rank with a cylindrical aperture.
The radius of the cylindrical aperture depends on the stellar mass of the galaxy and the depth of the cylinder is taken to be $\pm$ 2000~\kms. 
A galaxy with a stellar mass rank of 1 is identified as a central galaxy, whereas a mass rank greater than 1 is associated with a satellite galaxy. 
See \cite{fossati2015} and \cite{arora2019} for a description of the stellar mass rank scheme to identify centrals and satellites.
We have also cross-correlated the SDSS and MaNGA catalogues to compute the probability a galaxy being central or satellite according to the scheme of \cite{fossati2017}.

\begin{figure*}
    \centering
    \includegraphics[width=\linewidth]{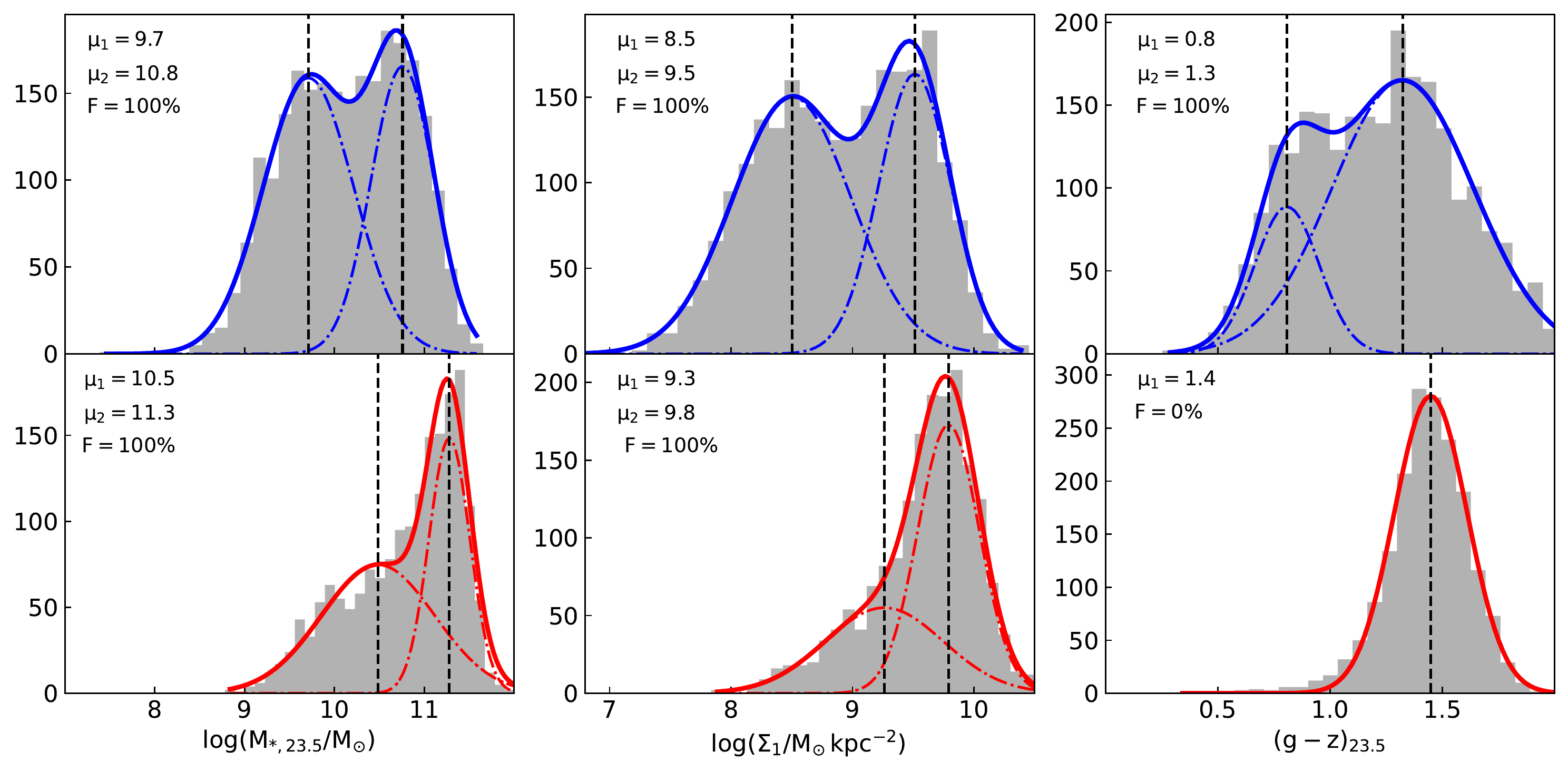}
    \includegraphics[width=\linewidth]{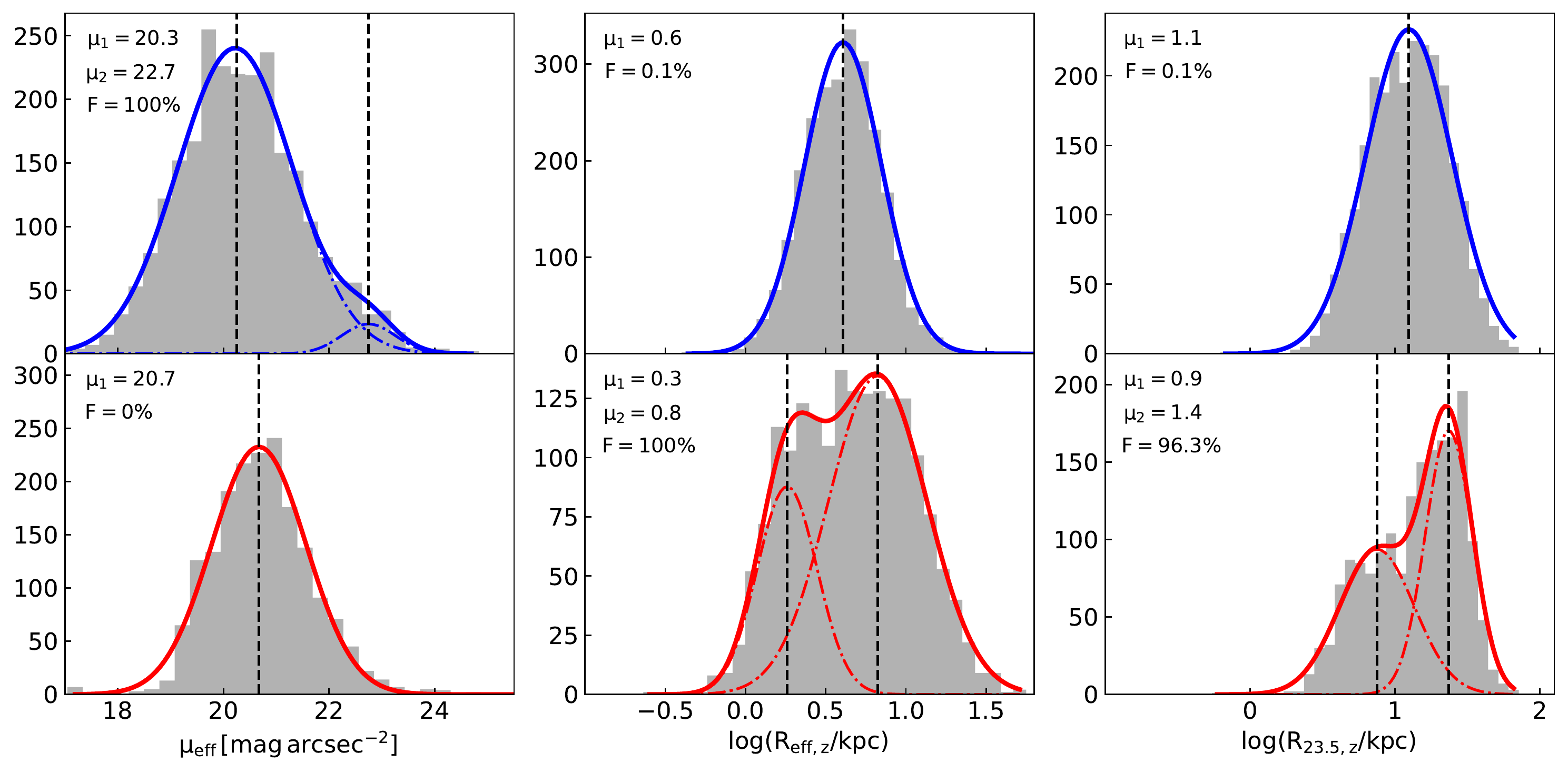}
    \caption{Distribution of MaNGA galaxy properties separated by galaxy morphology; LTGs in blue and ETGs in red. Shown are the stellar mass ($M_{*, 23.5}$), stellar mass surface density within 1~kpc ($\Sigma_{1}$), colour measured within $R_{\rm 23.5}$ ($(g-z)_{23.5}$), effective surface brightness ($\mu_{\rm eff}$), effective radius ($R_{\rm eff}$), and isophotal radius at 23.5~\magss ($R_{\rm 23.5}$). The grey histogram in each panel shows the underlying data distribution. Each LTG/ETG property is fit with a double or single Gaussian; the optimal distribution is determined via the F-test (see the text for details). The data in each panel give the Gaussian mean(s) of the unimodal (bimodal) distributions and the F-test confidence for a single vs. double Gaussian distribution.}
    \label{fig:manga_bimodal}
\end{figure*}

\subsection{Parameter corrections}

Our apparent magnitudes are corrected for Galactic extinction ($A_{\rm g}$), geometric projections ($\gamma$), internal extinction ($A_{\rm i}$), and k-correction ($A_{\rm k}$). 
Using the following transformation in each band:
\be
    m_{{\rm corr}, \lambda} = m_{{\rm obs}, \lambda} - A_{{\rm g},\lambda} - A_{\gamma, \lambda} - A_{{\rm k}, \lambda},
    \label{eq:corr}
\ee
\noindent the Galactic extinction ($A_{\rm g}$) is obtained from the NSA-Sloan Atlas\footnote{http://www.nsatlas.org/}
\cite[NSA;][]{Blanton2011}, which are themselves taken from \citet{Schlegel1998}.
The cosmological k-correction ($A_{\rm k}$) uses the template from \cite{Blanton2007}.

Correcting for internal extinction and geometric projections is more challenging, and depends on morphological type.  
Note that the discussion below applies to late-type galaxies (LTGs); ETGs are assumed dust-free and trixial and no such corrections are applied.  
MaNGA LTGs are identified via the MaNGA Deep Learning Morphological VAC (MDLM-VAC; \citealp{mldl}). 
The effects of projection on the sky bias the measurement of intrinsic properties, hence our attempt to recover face-on equivalent measures with a simple model \citep{Holmberg1958, Stone2020}. 
A common correction for the deprojection of structural parameters involves a linear fit between the desired projected variable and the log of the cosine of the inclination~\citep{Giovanelli1994}.
This method is expressed in \Eq{icorr}:
\be
    \log X_0 = \log X_i + \gamma\log(\cos(i)),
    \label{eq:icorr}
\ee
where $X_0$ is the variable corrected to face-on, $X_i$ is the observed measurement, $i$ is the inclination of the galaxy on the sky corrected for the stellar disc thickness and calculated using \eq{incl}, and $\gamma$ is the fitted correction factor:

\be
    \cos^2i = \frac{(b/a)^2-q_0^2}{1-q_0^2}; \quad (q_0=0.13).
    \label{eq:incl}
\ee

These corrections can be performed on a complete sample \citep{Masters2010} or applied to subsamples based on galaxy properties for which inclination corrections are robust and known such as \hi\ line widths \citep{Tully1998}, galaxy morphology \citep{Maller2009,Masters2010}, or total infrared luminosity \citep{Devour2019}.
The correction factor, $\gamma$, encompasses effects due to geometry (line-of-sight projection), stellar populations, and dust extinction as a function of inclination.

\begin{table*}
\centering
\begin{tabular}{ccccccccccc}
\toprule
T-Type & N    & $R_{23.5, g}$ & $R_{23.5, r}$ & $R_{23.5, z}$ & $L_{23.5, g}$  & $L_{23.5, r}$     & $L_{23.5, z}$  & $(g-r)_{23.5}$ & $(g-z)_{23.5}$ & $\log \Sigma_1$ \\
(1)    & (2)  & (3)           & (4)           & (5)           & (6)            & (7)               & (8)            & (9)            & (10)           & (11)            \\    \midrule
0--10   & 1907 & $0.28\pm0.03$ & $0.25\pm0.03$ & $0.23\pm0.03$ & $-0.34\pm0.08$ & $-0.24\pm0.09$   & $-0.12\pm0.09$  & $0.22\pm0.02$  & $0.51\pm0.04$  & $\phantom{-}0.01\pm0.08$  \\
0--3    & 829  & $0.38\pm0.05$ & $0.37\pm0.05$ & $0.35\pm0.05$ & $-0.05\pm0.11$ & $\phantom{-}0.10\pm0.12$  & $\phantom{-}0.25\pm0.12$  & $0.34\pm0.03$  & $0.72\pm0.04$  & $\phantom{-}0.43\pm0.11$   \\
4--5    & 958  & $0.21\pm0.04$ & $0.19\pm0.04$ & $0.18\pm0.04$ & $-0.42\pm0.11$ & $-0.28\pm0.11$    & $-0.15\pm0.11$ & $0.26\pm0.02$  & $0.62\pm0.04$  & $\phantom{-}0.05\pm0.11$   \\
6--10   & 164  & $0.31\pm0.08$ & $0.26\pm0.08$ & $0.22\pm0.08$ & $-0.28\pm0.23$ & $-0.28\pm0.24$    & $-0.08\pm0.24$ & $0.01\pm0.03$  & $0.09\pm0.08$  & $-0.25\pm0.18$   \\ \bottomrule
\end{tabular}
\caption{Inclination correction terms, $\gamma$, for various galaxy structural parameters in different photometry bands. Column (1) shows the T-Type bins used for the correction; column (2) presents the number of data points used for the fitting; and columns (3)--(11) show the correction factors for different structural properties. The first row gives $\gamma$ for all galaxies; the next three rows give $\gamma$ for specific morphological bins.}
\label{tab:gamma_corr}
\end{table*}
 
We solve \Eq{icorr} for $\gamma$ with a forward least-squares fit between the desired variable and $\cos(i)$ in log space.
For the fitting, we restrict the sample to an inclination range of $30^{\circ}<i<80^{\circ}$ and assume an intrinsic thickness of $q_0 = 0.13$ to convert the ellipticity to inclination \citep{Hall2012, Ouellette2017}.
Errors on $\gamma$ are calculated using bootstrap sampling.
\tab{gamma_corr} tabulates the correction factor, $\gamma$, for various galaxy structural properties of the MaNGA LTG sample in the {\it grz} bands. We present $\gamma$ for the full sample (first row of \tab{gamma_corr}) and split into three morphological bins done using the MLDM-VAC (last three rows of \tab{gamma_corr}).

The absolute value of these correction factors differ from those presented in \citet{Masters2010} and \citet{Stone2020}. 
However, in all cases, the correction factors are small.
For example, the correction factor for an isophotal radius in the transparent case is $\gamma\approx 2.3$~\citep{Giovanelli1994}, compared to our correction factor of $\rm {\sim}0.25\pm0.02$.

\begin{table*}
\begin{tabular}{@{}ccccccc@{}}
\toprule
Scaling relation          & \multicolumn{2}{c}{Uncorrected} & \multicolumn{2}{c}{Corrected T--Types 1-10} & \multicolumn{2}{c}{Corrected 3 T-Type bins} \\ 
                          & m               & $\sigma$       & m                 & $\sigma$          & m                 & $\sigma$          \\
(1)                       & (2)             & (3)            & (4)               & (5)               & (6)               & (7)               \\ \midrule
$\Sigma_1$--$M_{\rm *,23.5}$     & $0.96\pm0.01$   & $0.24\pm0.01$  & $0.96\pm0.01$     & $0.23\pm0.01$     & $0.96\pm0.01$     & $0.23\pm0.01$     \\
$R_{23.5, z}$--$M_{*,23.5}$  & $0.34\pm0.01$   & $0.11\pm0.01$  & $0.34\pm0.01$     & $0.11\pm0.01$     & $0.35\pm0.01$     & $0.10\pm0.01$     \\
$R_{23.5,z}$--$\Sigma_{1}$   & $0.30\pm0.01$   & $0.17\pm0.01$  & $0.29\pm0.01$     & $0.16\pm0.01$     & $0.30\pm0.01$     & $0.17\pm0.01$     \\
$R_{23.5, z}$--$L_{23.5,z}$  & $0.41\pm0.01$   & $0.08\pm0.01$  & $0.41\pm0.01$     & $0.09\pm0.01$     & $0.41\pm0.01$     & $0.09\pm0.01$     \\
$\Sigma_{1}$--$L_{23.5,z}$   & $0.94\pm0.01$   & $0.30\pm0.01$  & $0.95\pm0.01$     & $0.30\pm0.01$     & $0.93\pm0.01$     & $0.29\pm0.01$     \\
$(g-z)_{23.5}$--$L_{23.5,z}$ & $0.24\pm0.01$   & $0.24\pm0.01$  & $0.25\pm0.01$     & $0.22\pm0.01$     & $0.22\pm0.01$     & $0.21\pm0.01$     \\ \bottomrule
\end{tabular}
\caption{Variation of slopes and scatters of galaxy structural scaling relations due to inclination corrections. Column (1) indicates the scaling relation; columns (2) and (3) give the uncorrected slope and scatter; columns (4) and (5) give the slope and scatter with a single full-sample correction (all T-Types); columns (6) and (7) give the slopes and scatter with corrections applied to three morphological bins.}
\label{tab:gamma_sr}
\end{table*}

Ideally, an accurate inclination correction model should yield a reduction of the observed scatter in various galaxy scaling relations. 
This concept is tested in \tab{gamma_sr}, which presents the variation of the slopes and scatter for a suite of structural galaxy scaling relations with inclination corrections.
The slopes and scatter are found to be robust against inclination corrections for nearly all scaling relations presented. 
A significant variation in the scatter of the $(g-z)_{23.5}--L_{23.5,z}$ relation is found for both inclination correction models, with or without binning by morphology. 
In light of the null variation in the slopes and scatter of galaxy scaling relations, we apply inclination corrections to our photometry without reducing our sample into various morphological bins.

\begin{table*}
\centering
\begin{tabular}{@{}llcc@{}}
\hline
Column name            & Description                                                    & Unit                   & Data yype \\ \hline
MANGA-ID               & MaNGA Identification                                           & --                    & string    \\
PlateIFU               & MaNGA Plate-IFU                                                & --                    & string    \\
ObjID                  & SDSS-DR15 photometric identification number                    & --                    & long int  \\
RA                     & Object Right Ascension (J2000)                                 & $^\circ$                & float     \\
DEC                    & Object Declination (J2000)                                     & $^\circ$                & float     \\
Z                      & NSA or SDSS redshift                                           & --                    & float     \\
TTYPE                  & Morphological T-Type (from MDLM-VAC)                           & --                    & float     \\
GAL\_EXTINCTION        & Galactic extinction (DESI {\it grz} only)                      & mag                    & float     \\
KCORRECTION            & Cosmological {\it K-Corrections} (DESI {\it grz} only)               & mag                    & float     \\
R20                    & Radius where 20 per cent of the total light is integrated             & arcsec                 & float     \\
\texttt{}{R20\_E}      & Error in R20                                                   & arcsec                 & float     \\
Reff                   & Effective radius                                               & arcsec                 & float     \\
\texttt{}{Reff\_E}     & Error in effective radius                                      & arcsec                 & float     \\
R80                    & Radius where 80 per cent of the total light is integrated             & arcsec                 & float     \\
\texttt{}{R80\_E}      & Error in R80                                                   & arcsec                 & float     \\
R22.5                  & Isophotal radius calculated at $\rm 22.5\,\magss$              & arcsec                 & float     \\
\texttt{}{R22.5\_E}    & Error in R22.5                                                 & arcsec                 & float     \\
\texttt{}{R22.5\_FLAG} & Method of calculation: interpolation (0) and extrapolation (1) & boolean                & int     \\
R23.5                  & Isophotal radius calculated at $\rm 23.5\,\magss$              & arcsec                 & float     \\
\texttt{}{R23.5\_E}    & Error in R23.5                                                 & arcsec                 & float     \\
\texttt{}{R23.5\_FLAG} & Method of calculation: interpolation (0) and extrapolation (1) & boolean                & int     \\
R25                    & Isophotal radius calculated at $\rm 25\,\magss$                & arcsec                 & float     \\
\texttt{}{R25\_E}      & Error in R25                                                   & arcsec                 & float     \\
\texttt{}{R25\_FLAG}   & Method of calculation: interpolation (0) and extrapolation (1) & boolean                & int     \\
C25                    & Concentration index measured using R20 and Reff                & --                    & float     \\
C28                    & Concentration index measured using R20 and R80                 & --                    & float     \\
\texttt{}{MU\_20}      & Surface brightness at R20                                      & $\magss$               & float     \\
\texttt{}{MU\_20\_E}   & Error in \texttt{}{Mu\_20}                                     & $\magss$               & float     \\
\texttt{}{MU\_20\_FLAG}& Method of calculation: interpolation (0) and extrapolation (1) & boolean                & int     \\
\texttt{}{MU\_EFF}     & Surface brightness at the effective radius                     & $\magss$                 & float     \\
\texttt{}{MU\_EFF\_E}  & Error in effective surface brightness                          & $\magss$                 & float     \\
\texttt{}{Mu\_EFF\_FLAG} & Method of calculation: interpolation (0) and extrapolation (1) & boolean              & int     \\
\texttt{}{MU\_80}      & Surface brightness at R80                                      & $\magss$                 & float     \\
\texttt{}{MU\_80\_FLAG}& Error in \texttt{}{Mu\_80}                                     & $\magss$                 & float     \\
\texttt{}{MU\_80\_FLAG}& Method of calculation: interpolation (0) and extrapolation (1) & boolean              & int     \\
MAG23.5                & Total apparent magnitude within R23.5                          & mag                    & float   \\
\texttt{}{MAG23.5\_E}    & Error in total apparent magnitude within R23.5                 & mag                    & float   \\
\texttt{}{MAG23.5\_FLAG} & Method of calculation: interpolation (0) and extrapolation (1) & boolean                & int   \\
MAG25                  & Total apparent magnitude within R25                            & mag                    & float   \\
\texttt{}{MAG25\_E}      & Error in total apparent magnitude within R25                 & mag                    & float   \\
\texttt{}{MAG25\_FLAG}   & Method of calculation: interpolation (0) and extrapolation (1)& boolean               & int   \\
L23.5                  & Total luminosity within R23.5                                  & $L_{\odot}$           & float     \\
L25                    & Total luminosity within R25                                    & $L_{\odot}$           & float     \\
ELLIPTICITY            & Ellipticity of the last isophote; $1-(b/a)$                    & --                    & float    \\
PA                     & Position angle of the last isophote measured from north to east (DESI {\it grz} only)& $^\circ$           & float  \\
MSTAR\_235             & Stellar mass measured at the {\it z}-band R$_{23.5}$ ({\it z} \& {\it W}1 band only) & $\rm M_{\odot}$  & float\\
MSTAR\_235\_E          & Error in stellar mass estimates measured at {\it z} band R$_{23.5}$ ({\it z} band only)     & $\rm M_{\odot}$& float\\
MSTAR\_25              & Stellar Mass measured at the {\it z}-band R$_{25}$ (z band only)    & $\rm M_{\odot}$        & float\\
MSTAR\_25\_E           & Error in stellar mass estimates measured at {\it z}-band R$_{25}$ ({\it z} band only)    & $\rm M_{\odot}$ & float\\
\texttt{}{SIGMA1}      & Stellar mass surface density within 1\,kpc  (z band only)      & $\rm M_{\odot}\,kpc^{-2}$ & float     \\
\texttt{}{SIGMA1\_E}   & Error in SIGMA1  (z band only)                                 & $\rm M_{\odot}\,kpc^{-2}$ & float     \\
\texttt{}{SIGMA\_EFF}  & Stellar mass surface density within R$_{\rm eff}$  (z band only)        & $\rm M_{\odot}\,kpc^{-2}$ & float     \\
\texttt{}{SIGMA1\_EFF\_E}& Error in SIGMA\_EFF  (z band only)                                 & $\rm M_{\odot}\,kpc^{-2}$ & float     \\
COLOR                  & $g-r$ and $g-z$ measured at z-band R$_{23.5}$ and R$_{25}$ (z band only) & mag  & float     \\
\hline
\end{tabular}
\caption{Photometric quantities for the MaNGA DESI photometric catalogue. The table includes calculated parameters along with the units and the data types. All parameters presented are uncorrected for Galactic extinction, inclination and cosmology.
These parameters are available in three separate files for the DESI \textit{grz} bands. 
The three full tables are presented as supplementary material.}
\label{tab:phototable}
\end{table*}

\begin{table*}
\centering
\begin{tabular}{@{}llcc@{}}
\hline
Column Name          & Description                                                                                & Unit    & Data type \\ \hline
MANGA-ID             & MaNGA Identification                                                                       & --     & string    \\
PlateIFU             & MaNGA Plate-IFU                                                                            & --     & string    \\
ObjID                & SDSS-DR15 photometric identification number                                                & --     & long int  \\
RA                   & Object Right Ascension (J2000)                                                             & $^{\circ}$ & float     \\
DEC                  & Object Declination (J2000)                                                                 & $^{\circ}$ & float     \\
Z                    & NSA or SDSS redshift                                                                       & count     & float     \\
\texttt{}{Dens\_01}  & Number of neighbours within a 0.1-Mpc aperture                                             & count     & int       \\
\texttt{}{Dens\_02}  & Number of neighbours within a 0.2-Mpc aperture                                             & count     & int       \\
\texttt{}{Dens\_05}  & Number of neighbours within a 0.5-Mpc aperture                                             & count     & int       \\
\texttt{}{Dens\_1}   & Number of neighbours within a 1-Mpc aperture                                               & count     & int       \\
\texttt{}{Dens\_2}   & Number of neighbours within a 2-Mpc aperture                                               & count     & int       \\
\texttt{}{Dens\_3}   & Number of neighbours within a 3-Mpc aperture                                               & count    & int       \\
\texttt{}{Mrank\_AA} & Identification of central or satellite galaxy; central (1) and satellite (\textgreater{}1) & rank     & int       \\
\texttt{}{P\_CEN}    & Probability of the galaxy being the central member in the halo                                  & rank    & float     \\
\texttt{}{P\_SAT}    & Probability of the galaxy being a satellite in the halo                                   & --     & float     \\ \hline
\end{tabular}
\caption{Environmental properties for the MaNGA galaxies. The table includes each parameter along with the units and data types.
The full table is presented as supplementary material.}
\label{tab:manga_env}
\end{table*}

\section{Parameters Distributions} 
\label{sec:param_dist}

This section provides an overview of some important MaNGA galaxy properties and their range.
\fig{manga_bimodal} offers a broad appreciation of the distribution of structural properties for galaxies in the MaNGA public release. 
For instance, if dwarf galaxies are defined as having a stellar mass $\log(M_*/M_{\odot}) \leq 9.5$ \citep{Woo2008, Ouellette2017}, only 11 percent of our galaxies are dwarfs. 
Our results in this section clearly highlight the presence of distinct populations in each of the LTG and ETG categories. 
The latter is made clear through the unimodal and bimodal trends displayed by some structural and dynamical parameters \citep{Tully1997,McDonald2009,Sorce2013,Ouellette2017}.
We define a unimodal distribution as one described by a normal distribution.  
The figure of merit to distinguish a bimodal from a unimodal distribution is the F-test that gives the probability that the observed distributions originate from two distinct Gaussian populations rather than a single one. 
These populations are represented by single or double Gaussian fits in \fig{manga_bimodal}. 
F-test confidence results for a double vs a single Gaussian distribution and the fitted means for the double/single population(s) are indicated in each panel. 
Unimodal distributions are found for LTG sizes (effective and isophotal) and ETG effective surface brightnesses and colour.

The following properties for LTGs show bimodal signatures; stellar mass (top left-hand panel), stellar surface density with 1~kpc (middle panel), effective surface brightness (bottom left-hand) and colour (top right-hand) measured at 23.5~\magss.
The notion of surface brightness bimodalities has been addressed at some length in \citep[][and references therein]{Ouellette2017}. 
Surface brightness bimodalities in LTGs are especially conspicuous at dust-insensitive wavelengths such as the {\it z} band \citep{Sorce2013}.

For ETGs, stellar mass, stellar surface density within 1~kpc, isophotal radius at 23.5~\magss, and effective radius show bimodal signatures.
A bimodality in the galaxian properties shown here reflects environmental influences, varying merger and star formation histories, and different radial dark matter fractions, which all play a key role in galaxy evolution.
The coupling of these structural bimodalities with bimodal dynamical properties (if any) would greatly enhance galaxy formation scenarios.

The detailed investigation of bimodal trends in scaling relations, and the identification of galaxy subclasses, will be presented elsewhere. 
For now, we appreciate the need to consider one or two populations in fitting scaling relations over a range of galaxian properties. 

\section{Literature Comparisons} 
\label{sec:compare}
The following section presents comparisons of our galaxy structural parameters with similar values found in the literature.
We first present a multifaceted comparison of the 1D surface brightness profiles generated using model-independent (e.g. \ap, {\scriptsize XVISTA}) and model-dependent (e.g. {\scriptsize GALFIT}, \pym) methods.
Effective (half-light), isophotal radii and apparent magnitudes from the MPP-VAC of \citetalias{Fischer2019} are also compared with our photometry.
This comparison recognizes that the MPP-VAC reports parameters from circularized light profiles. 
We caution that this operation is especially uncertain in light of unknown dust distributions, non-axisymmetric feature and disc thicknesses, and the vagaries of 2D image decompositions \citep{Gilhuly2018}. 
Total magnitudes, and effective radii, should be the least model-biased parameters from \citetalias{Fischer2019}, and hence our comparison below. 
Stellar masses are also compared with a PCA-based stellar mass estimation technique from \citet[][hereafter \citetalias{Pace2019}]{Pace2019}.
For the sake of comparison, parameters presented in this section are not corrected for Galactic extinction, inclination, and cosmology.

\subsection{Surface brightness profile comparisons}
\label{sec:sbprof}

\begin{figure}
    \centering
    \includegraphics[width=\columnwidth]{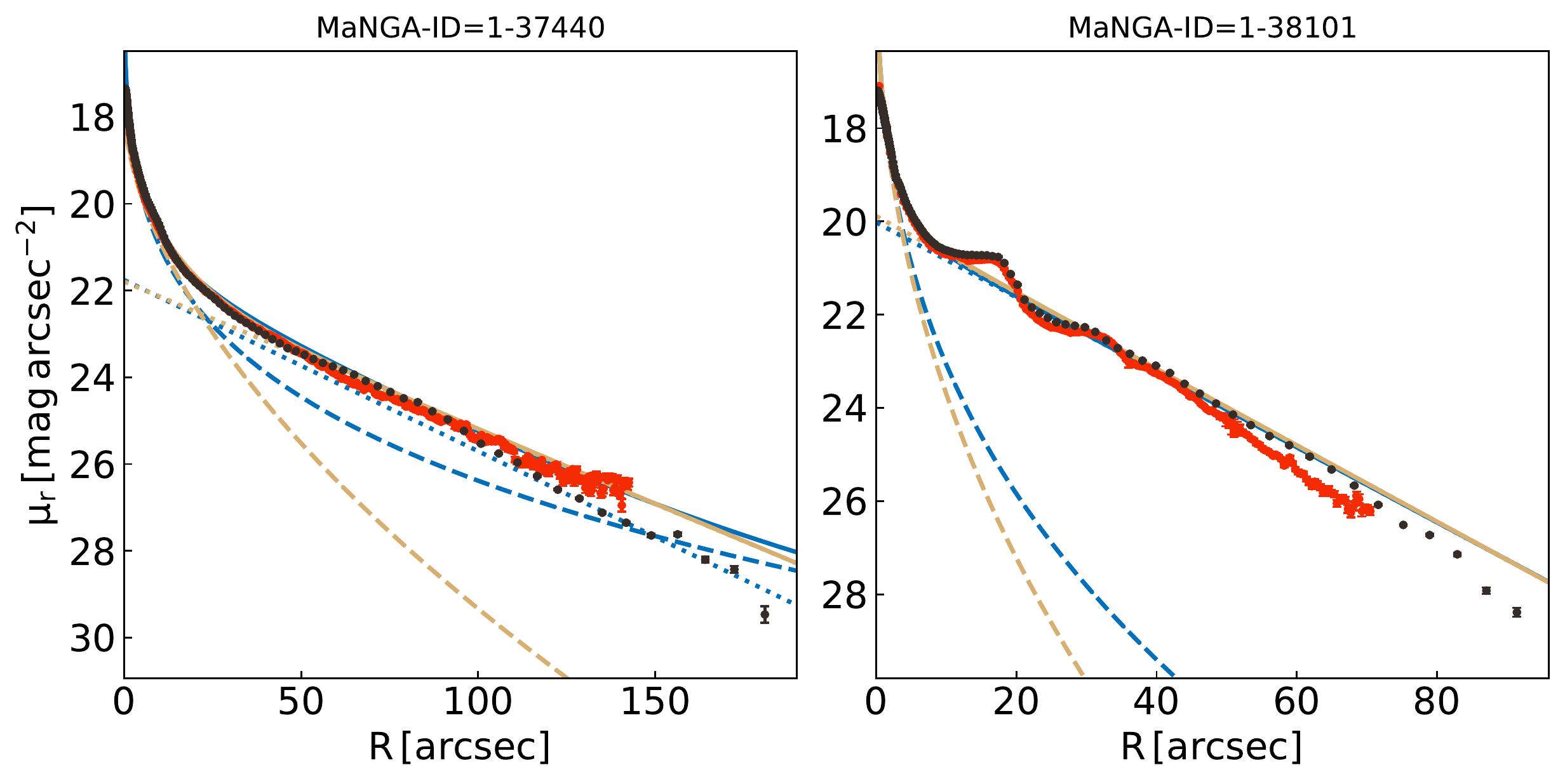}
    \includegraphics[width=\columnwidth]{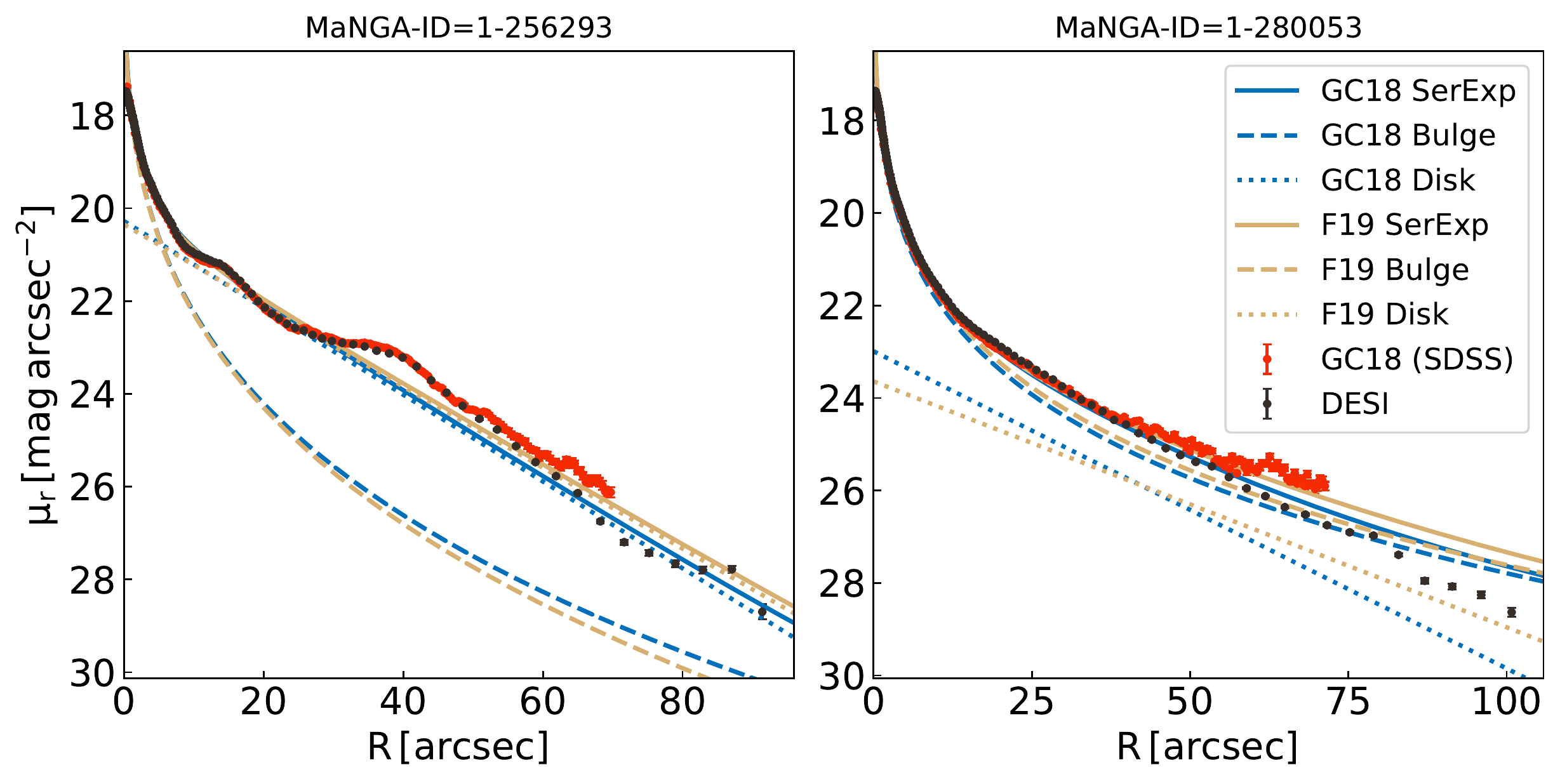}
    \includegraphics[width=\columnwidth]{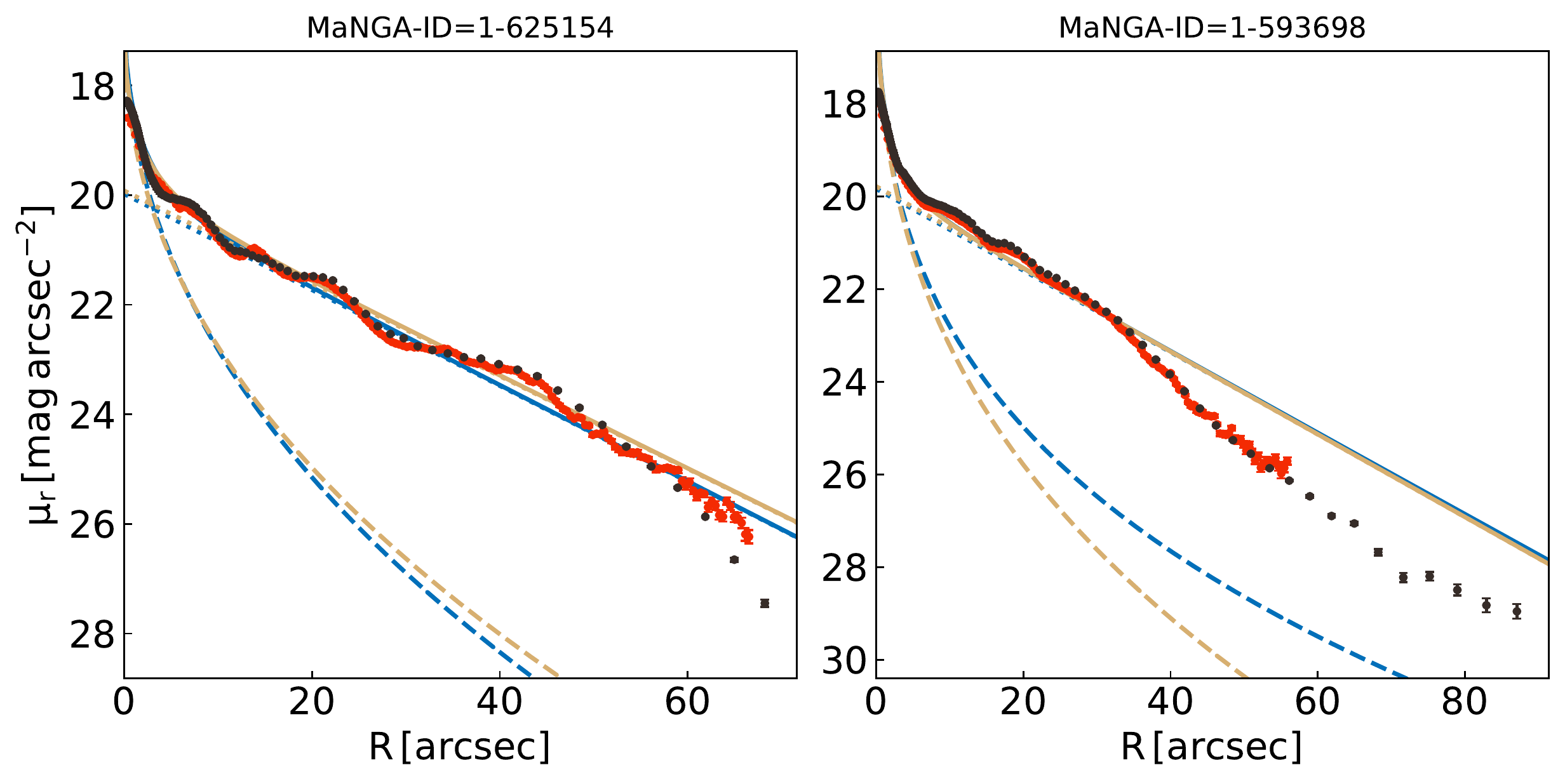}
     \includegraphics[width=\columnwidth]{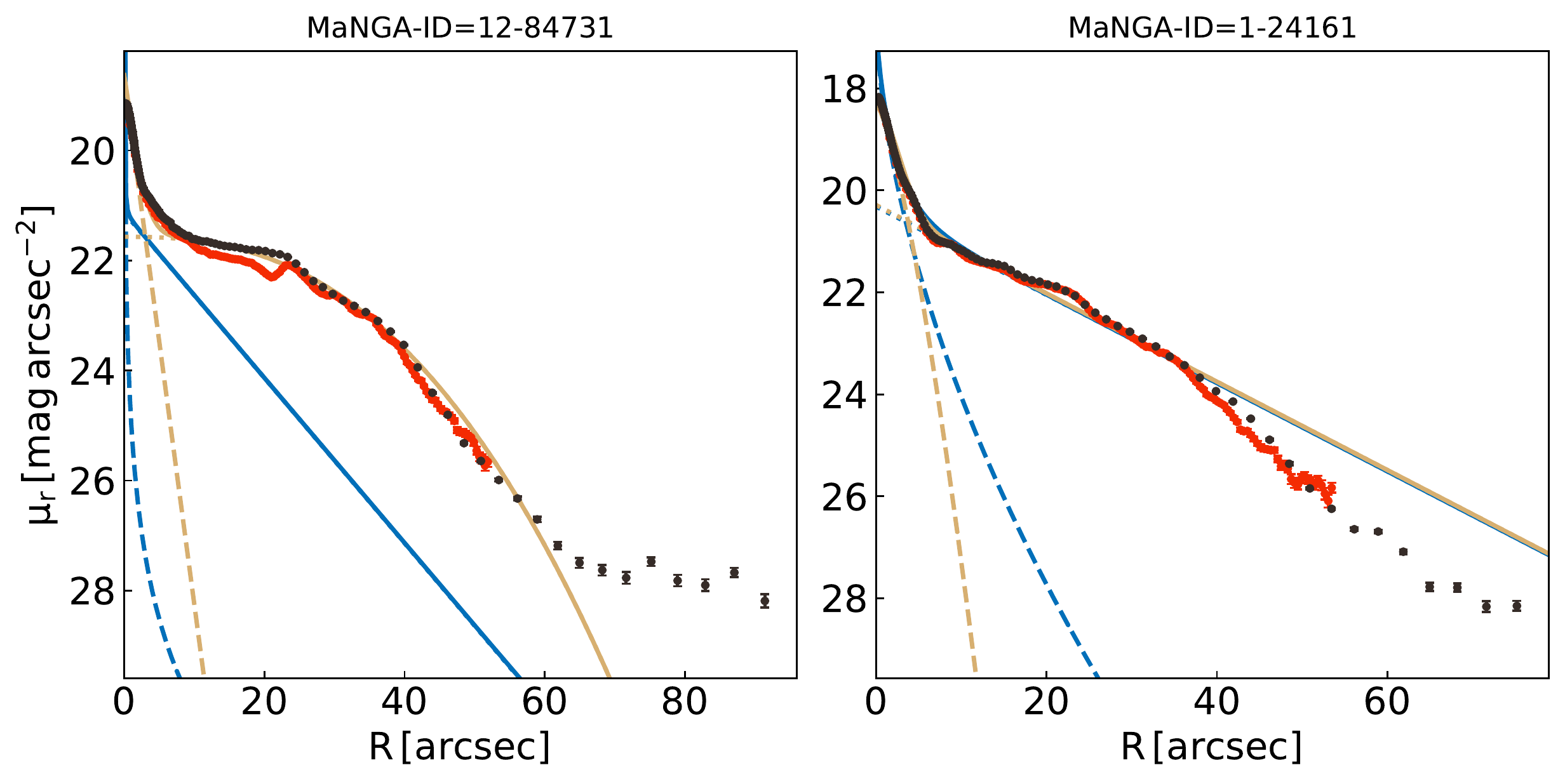}
    \caption{Comparison of our DESI-based 1D {\it r}-band surface brightness profile (black dots) with the SDSS-based surface brightness profile (red dot) of \citetalias{Gilhuly2018}. Also compared are the 2D S\'ersic + Exponential decompositions using {\scriptsize GALFIT} by \citetalias{Gilhuly2018} (cyan solid lines) and \citetalias{Fischer2019} (gold solid line). The S\'ersic + Exponential decompositions are separated by their respective bulge (dashed line) and disc (solid line) contributions. Missing coloured lines represent failed fits by the original authors.}
    \label{fig:sbcompare}
\end{figure}

\fig{sbcompare} compares 1D {\it r}-band surface brightness profiles extracted from DESI images (black points) by us and from SDSS images by \citetalias{Gilhuly2018} (red points).
The excellent agreement between those profiles results from quality DESI/SDSS imaging and the robust non-parametric surface brightness profile extraction with {\scriptsize XVISTA} and \ap. 
Figs. \ref{fig:vista_auto_compare} and \ref{fig:sbcompare} compare various global properties between {\scriptsize XVISTA} and \ap, showing that both codes reproduce local variations in 1D surface brightness profiles with high fidelity.

Our catalogue also overlaps with \citetalias{Fischer2019} and \citetalias{Gilhuly2018}.  
The latter two studies used the 2D image fitting algorithm {\scriptsize GALFIT} to obtain Bulge-Disc decompositions for their samples based on the MaNGA and CALIFA samples, respectively.  
A total of 16 galaxies overlap between the MaNGA (us and \citetalias{Fischer2019}) and CALIFA (\citetalias{Gilhuly2018}) samples.  
These overlapping galaxies enable a comparison of our independent results. 

In \fig{sbcompare}, the blue and gold solid line show the S\'ersic + Exponential decompositions by \citet{Gilhuly2018} and \citetalias{Fischer2019} respectively, for eight galaxies using {\scriptsize GALFIT}.
We caution that the MPP-VAC (\citetalias{Fischer2019}) present total apparent magnitudes in the circularized plane of the galaxy, i.e. corrected to face-on based on the simplest assumptions of a dust-free, infinitesimally-thin disc.
The recovery of total apparent magnitudes ($m$) from the MPP-VAC in the plane of the sky must therefore be made with the equation $(m_{r}' = m_{r} + 2.5\log(b/a))$ for the bulge and disc components, where $b/a$ is the axis ratio of the galaxy.

The appreciation of more complex disc systems, with their triaxial bulges, thicker mid-planes, and sporadic dust extinction, calls for a more extensive modelling. 
\cite{Stone2020} performed an extensive analysis of inclination corrections for disc galaxies.
While some correction models showed a scatter reduction of various galaxy scaling relations, little agreement between different correction models was found.

For six out of the eight galaxies, the two independent {\scriptsize GALFIT} decompositions return S\'ersic + Exponential models that can differ by as much as ${\sim}0.4\,\magss$, demonstrating the great subjectivity between these model-dependent solutions while our non-parametric comparison only show differences on the order of ${\sim}0.09\,\magss$.
In some cases, only one user can find a valid solution for the {\scriptsize GALFIT} decomposition.
Where both fits converge, they often find different bulge to disc ratios with a difference of ${\sim}0.3$\,dex.
Similar caveats for Bulge-Disc decompositions for the same galaxies using two different image fitters ({\scriptsize GALFIT} and {\scriptsize IMFIT} \citep{imfit}) by \citetalias{Gilhuly2018} showed discrepant results as well.  

The large variations between the solutions of \citetalias{Gilhuly2018} and \citetalias{Fischer2019}, as well as the analysis presented in \citetalias{Gilhuly2018}, remind us that generic parametric solutions are especially fragile and inconclusive. 
For these reasons, an analysis of galaxy structure or scaling relation should rely on the non-parametric characterization of surface brightness profiles. 

\subsection{Effective sizes}\label{sec:effsizes}
\begin{figure}
    \centering
    \includegraphics[width=\columnwidth]{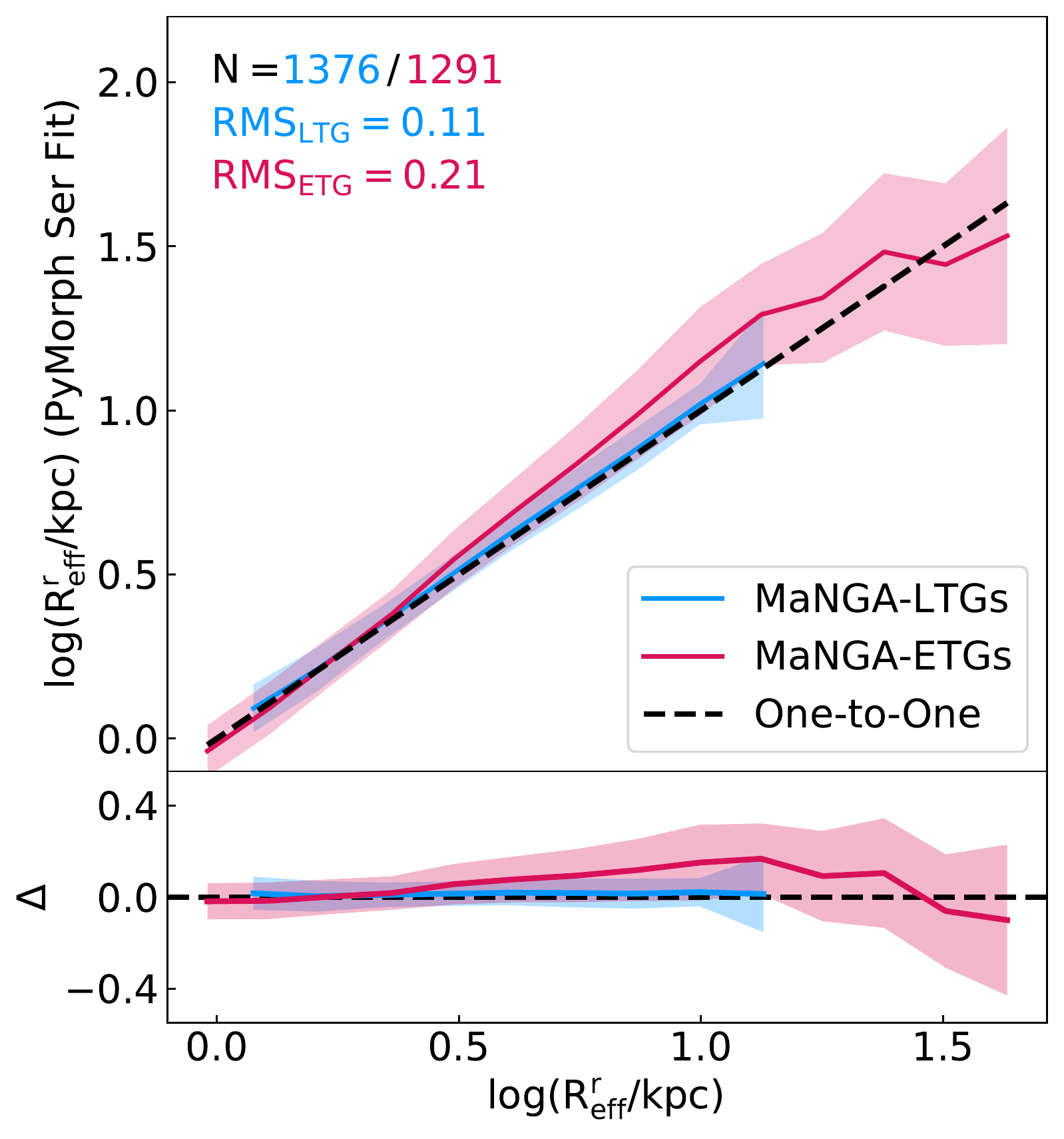}
    \caption{Comparison of our model-independent effective radii from DESI \textit{r}-band surface brightness profiles with those from a single S\'ersic fit from \citetalias{Fischer2019}. The solid lines show the median for MaNGA-LTGs (blue) and MaNGA-ETGs (red) and the shaded region shows the scatter within a bin of size 0.12 dex. The dashed line shows the one-to-one line. The inset text shows the number of data points and root mean square (rms) for the LTG and ETG populations.
    The bottom panel shows the residual $\Delta = \log (R^r_{\rm eff}[PYMORPH]/R^r_{\rm eff})$.}
    \label{fig:f18_size_ser}
\end{figure}

\begin{figure}
    \centering
    \includegraphics[width=\columnwidth]{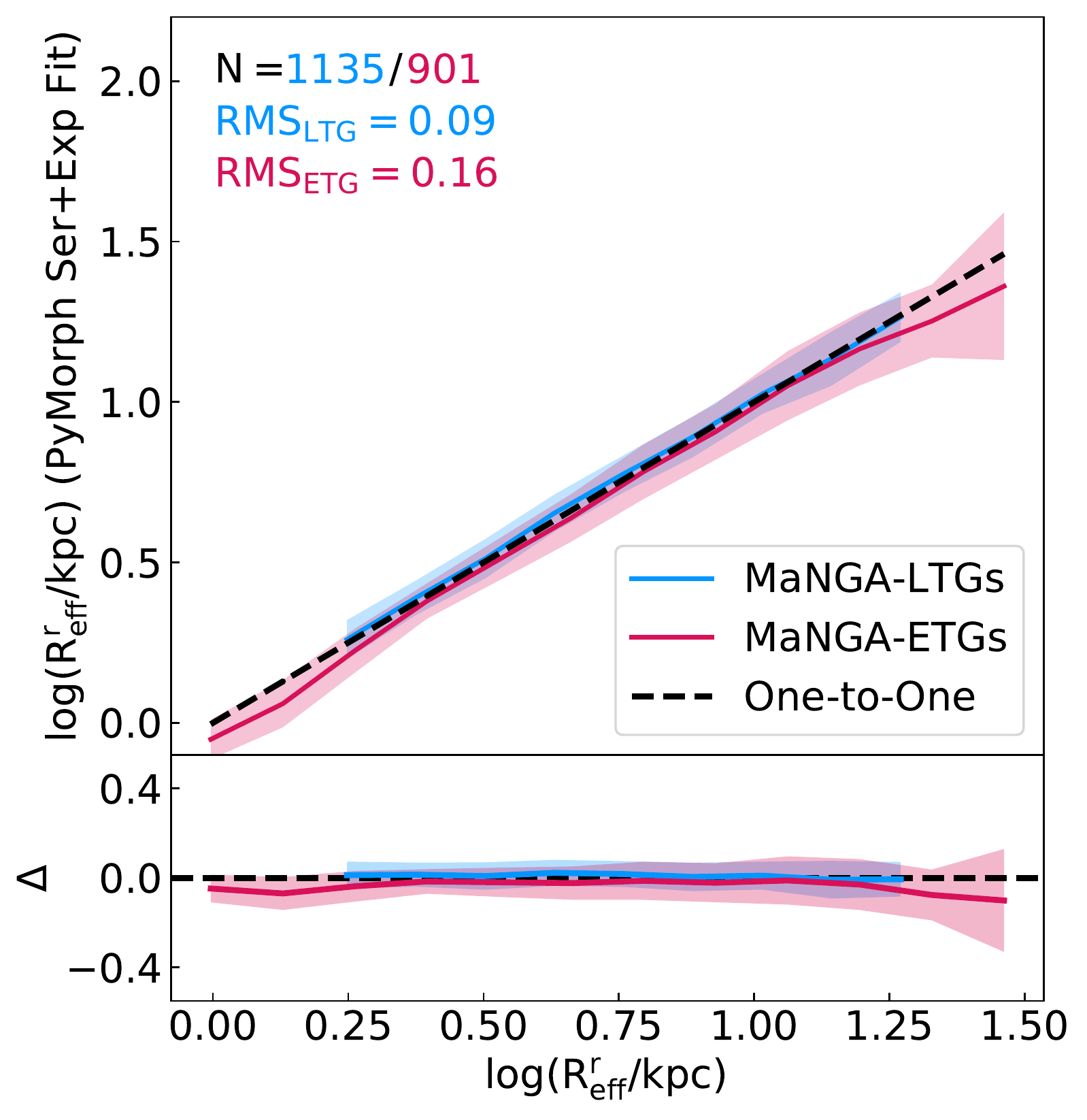}
    \caption{Same as \Fig{f18_size_ser} but the {\it y}-axis uses effective radii from S\'ersic and exponential fits by \citetalias{Fischer2019}.}
    \label{fig:f18_size_serexp}
\end{figure}

\Fig{f18_size_ser} compares our non-parametric $R_{\rm eff}$, measured in the DESI {\it r} band, with those found in \citetalias{Fischer2019} for a single S\'ersic fit with PyMorph.
We have used the MDLM-VAC to divide our sample into LTG and ETG categories. 
Size measurements are taken in the $r$-band to enable a direct comparison with PyMorph results. 
\pym and \ap agree well at small radii; at larger radii \ap yields smaller sizes than \pym.

This behaviour is expected as a single S\'ersic fit does not capture all the light from the outer regions of a galaxy, with low S/N, yielding a larger estimate for $R_{\rm eff}$.
The effect is amplified for larger \pym $R_{\rm eff}$ and for the ETG sample due to their higher light concentration relative to LTGs. 
For the complete population, a single S\'ersic fit from \citetalias{Fischer2019} results in larger effective radii by 0.21\,dex compared to our results.

The agreement between our methods improves with the two-component fits from \citetalias{Fischer2019}.
\Fig{f18_size_serexp} shows a comparison between our non-parametric effective radii and those extracted from two-component S\'ersic exponential fits. 
Both ETGs and LTGs are in agreement, though at large radii the scatter increases.

While the \pym two-component fit improves the overall agreement with \ap, there remains significant random variations, especially with quantities determined relative to the total light of the galaxy, like the effective radius, $R_{\rm eff}$.  
The latter suffered from poor reproducibility, largely due to the uncertain definition of total apparent magnitude of a galaxy.
In a similar vein, \citet{Hall2012} found that the scatter of 
velocity--radius--luminosity (VRL) scaling relations is reduced with isophotal radii and \citet{Trujillo2020} found that a stellar mass density radius reduces scatter in the size--mass relation.
Similar impressions were echoed by \citetalias{Gilhuly2018} who found differences as large as 0.16\,dex (45\%) between non-parametric and model-dependent measures of $R_{\rm eff}$ for CALIFA galaxies~\citep{Walcher2014}. 
\citet{Samaniego2017} also used the FIRE simulations to point out the same pathology about $R_{\rm eff}$.
For the size-dependent analyses that follow (\sec{sr}), we limit our use of effective radii and pay special attention to isophotal size metrics. 
The scatter in the size--mass relation is discussed in \sec{sr}.

\subsection{Isophotal sizes}

\begin{figure*}
    \centering
    \includegraphics[scale=0.48]{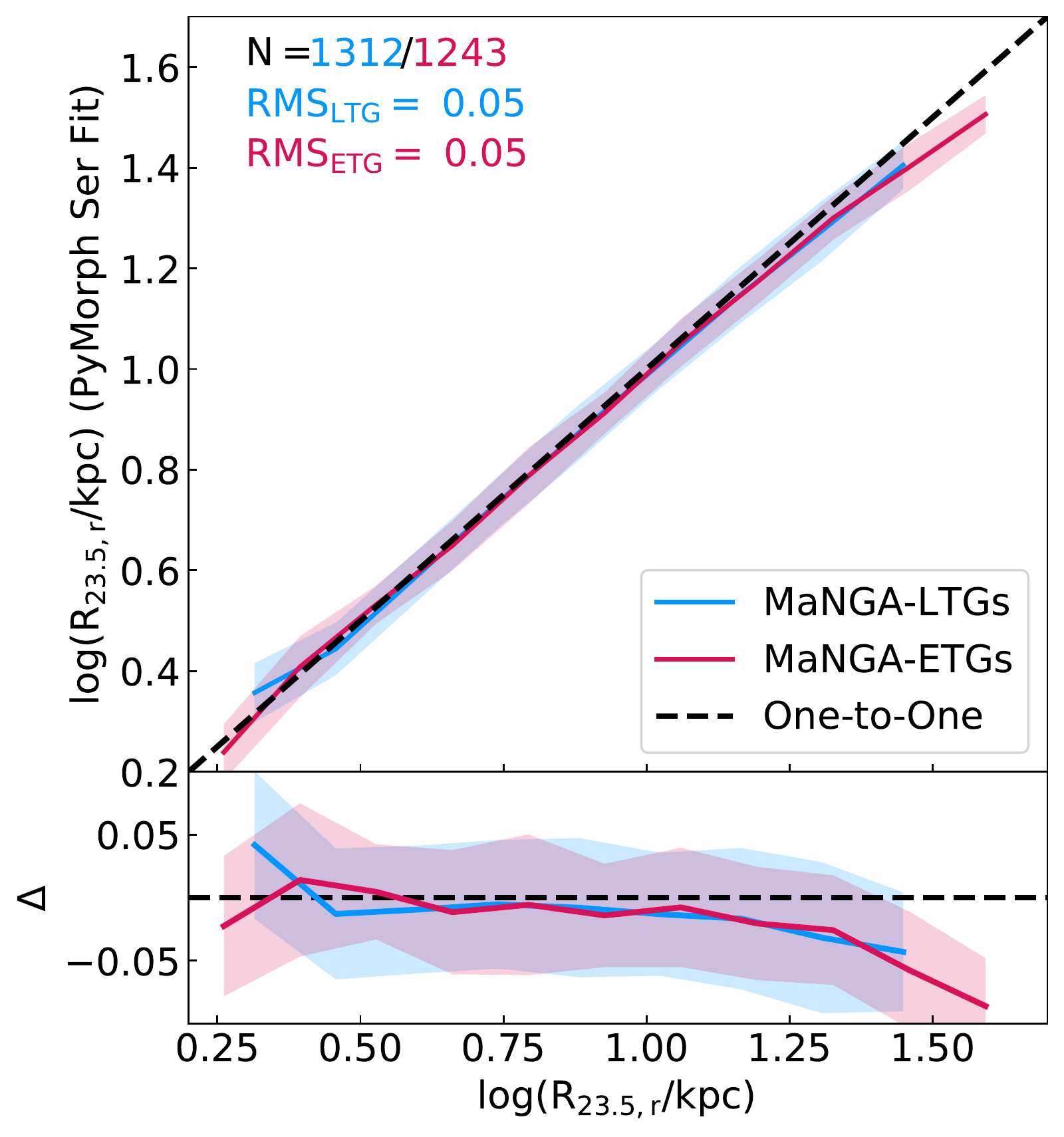}
    \includegraphics[scale=0.50]{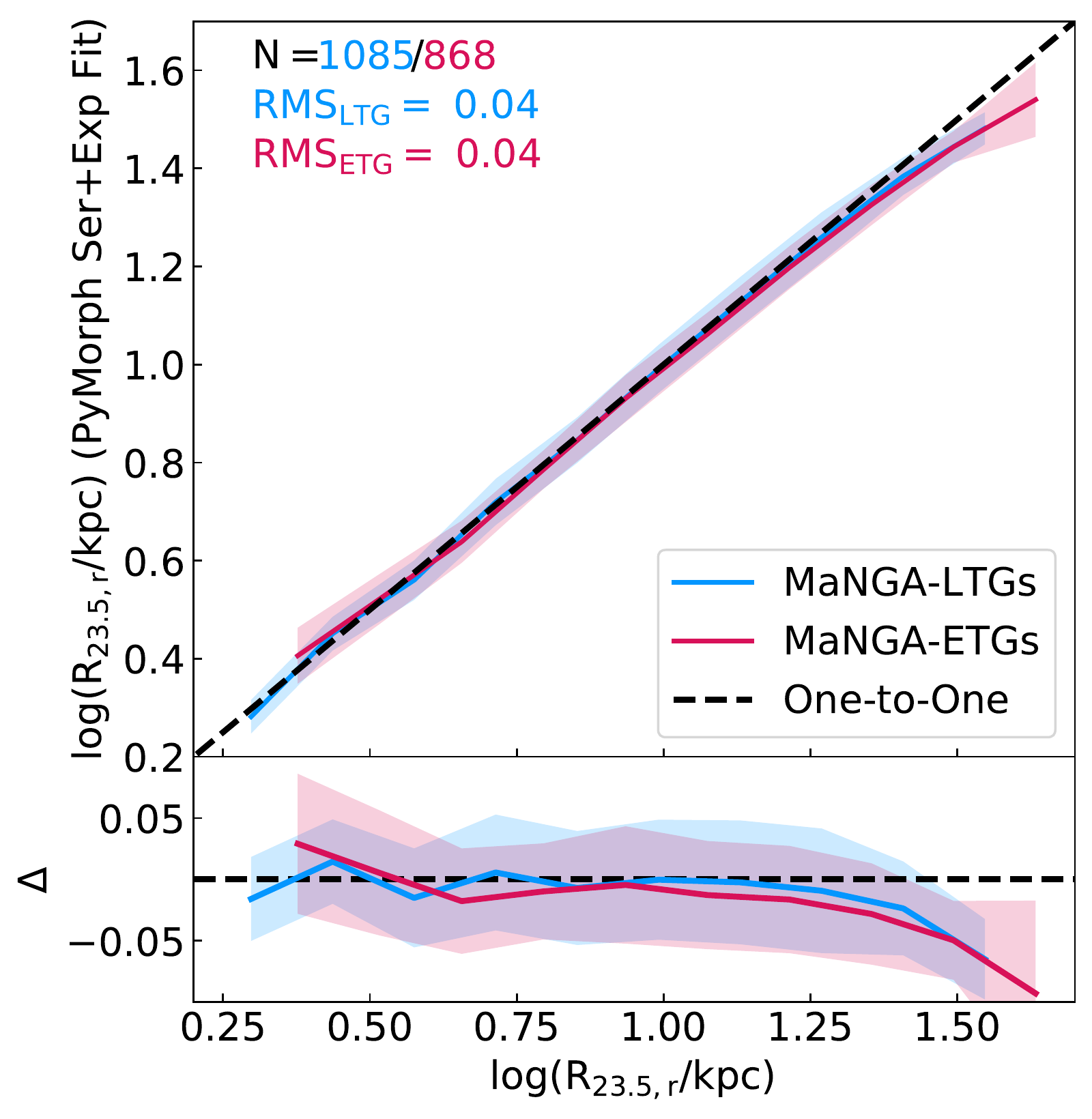}
    \caption{Comparison of our model-independent isophotal radius measured at $23.5$\,\magss from DESI \textit{r}-band surface brightness profiles with model-dependent \pym decomposition from \citetalias{Fischer2019}. The solid lines show the median for MaNGA-LTGs (blue) and MaNGA-ETGs (red), and the shaded region shows the scatter in a bin of width 0.12 dex. The dashed line is the one-to-one line. The legend shows the number of data points and rms for the LTG and ETG populations. The left- and right-hand panels show the S\'ersic and S\'ersic+Exponential fits, respectively. The bottom panels show the residual $\Delta = \log (R^r_{23.5}[PYMORPH]/R^r_{23.5})$.} 
    \label{fig:f18_r235}
\end{figure*}

The parameters from the MPP-VAC can be used to construct de-projected surface brightness profiles in order to infer the isophotal radii measured at $23.5$\,\magss.
These are compared with matching size measurements from the DESI photometry in \fig{f18_r235}.
Disagreements in isophotal sizes are larger for galaxies with preferred S\'ersic model (${\sim}0.05\,$dex) compared to galaxies with preferred S\'ersic+Exponential model   
(${\sim}0.04\,$dex).
This reaffirms the results from \sec{effsizes} that simple S\'ersic models cannot account for all of the galaxy light. 

The comparison of model-independent isophotal radii measured with {\scriptsize XVISTA} and \ap in \fig{vista_auto_compare} showed more consistency ($0.02\,$dex), indicating that non-parametric modelling of galaxies is more reproducible.  
The larger rms ($\sim$0.05\,dex) is explained by the vagaries in parametric modelling from GALFIT.
Comparing \fig{f18_r235} and \fig{f18_size_serexp} demonstrates that that isophotal sizes ($R_{\rm 23.5}$) are more consistent than sizes based on fraction of total light even while comparing non-parametric with model-dependent sizes.

\subsection{Apparent magnitudes}

\begin{figure}
    \centering
    \includegraphics[width=\columnwidth]{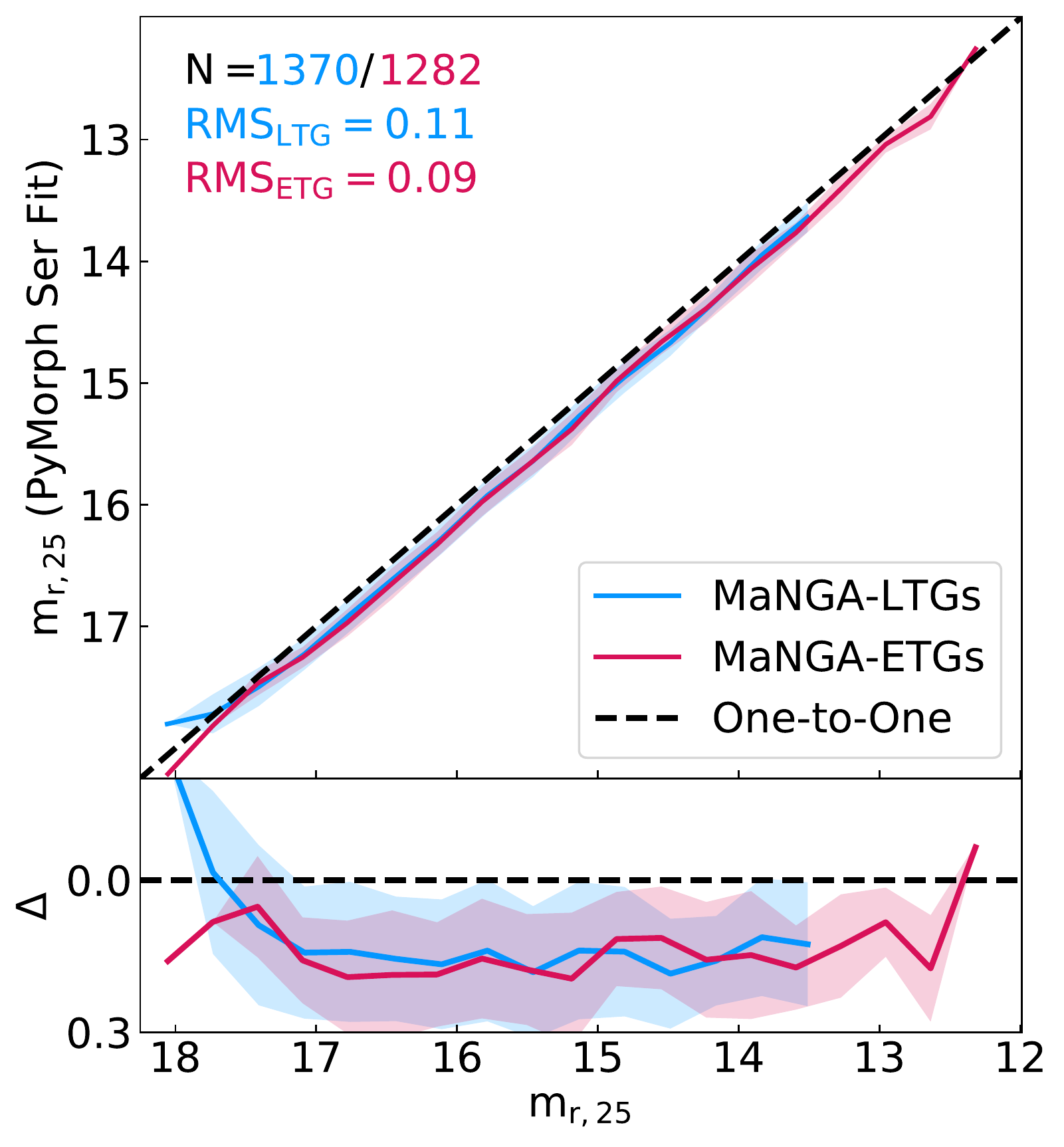}
    \caption{Comparison of our model-independent apparent magnitudes from DESI \textit{r}-band surface brightness profiles with those from a single S\'ersic fit from \citetalias{Fischer2019}. For both methods, the total apparent magnitude is measured within the $25\,\magss$ isophote. The bottom panel shows the residual.}
    \label{fig:f18_mag_ser}
\end{figure}

\begin{figure}
    \centering
    \includegraphics[width=\columnwidth]{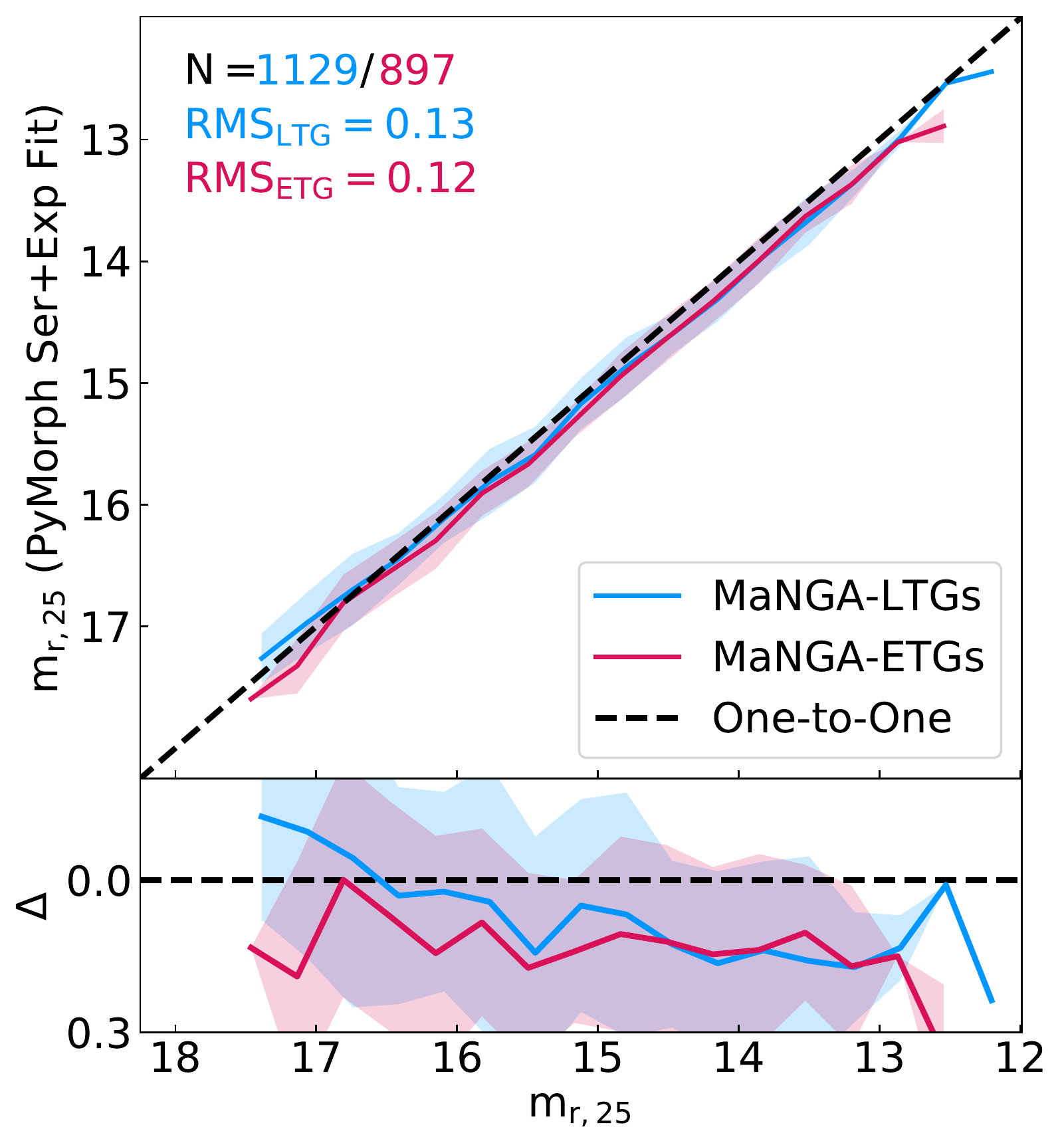}
    \caption{Same as \Fig{f18_mag_ser} with apparent magnitude on the y-axis for a double S\'ersic and exponential fit from \citetalias{Fischer2019}.}.
    \label{fig:f18_mag_serexp}
\end{figure}

Next, total apparent magnitudes extracted from DESI photometry are compared with MPP-VAC magnitudes from \citetalias{Fischer2019}.
The total apparent magnitudes are both calculated within $25\,\magss$ and are not corrected for cosmology, Galactic and inclination extinctions.
While total magnitudes are here mostly reported at the $23.5\,\magss$ levels to maximize comparisons with literature values, the current comparisons (\fig{f18_mag_ser} and \ref{fig:f18_mag_serexp}) at the $25\,\magss$ isophotal level takes advantage of the superior DESI imaging depth.
Once again, the \citetalias{Fischer2019} total apparent magnitudes are deprojected according to the expression, $m'=m+2.5\log(b/a)$.
\Fig{f18_mag_ser} compares apparent {\it r}-band magnitudes from our DESI surface brightness profiles with S\'ersic fit magnitudes from \citetalias{Fischer2019}, in various galaxy morphological bins.
The rms values are given in dex (magnitude divided by 2.5).
For both ETGs and LTGS, the DESI surface brightness profiles from \ap are $\sim 0.10$\,dex brighter than those from \pym with a S\'ersic fit.
This further highlights that a single S\'ersic fit fails to capture all the light from the object, leading to disagreements in effective galaxy sizes (\sec{effsizes}).

Surprisingly, the  addition of a second, exponential, component in \pym for MaNGA galaxies results in poorer fits (\fig{f18_mag_serexp}).
While random variations between \pym and \ap for LTGs are reduced with an additional component, PyMorph still underestimates the total light (calculated using the analytic S\'ersic function) relative to our analysis.
The shaded regions for both LTGs and ETGs in \fig{f18_mag_serexp} represent the inter-quartile range of the residuals.
With our large sample, we can estimate the random error (as $1/\sqrt{N}\sim 0.03\,$dex) to be very small, and the rms errors are thus largely systematic. 
Indeed, disagreements with our apparent magnitudes emerge largely from the model-dependent nature of the \citetalias{Fischer2019} photometric analysis that do not account for non-axisymmetric features such as bars, rings and spiral arms. 
Features unaccounted for by \pym will systematically yield fainter total magnitudes relative to non-parametric estimates. 
Along with the surface brightness profile comparison in \sec{sbprof}, the size and apparent magnitude comparisons further reinforce the benefits of using model-independent technique for measuring galaxy structural properties.

\subsection{Stellar masses}

\begin{figure*}
    \centering
    \includegraphics[width=\linewidth]{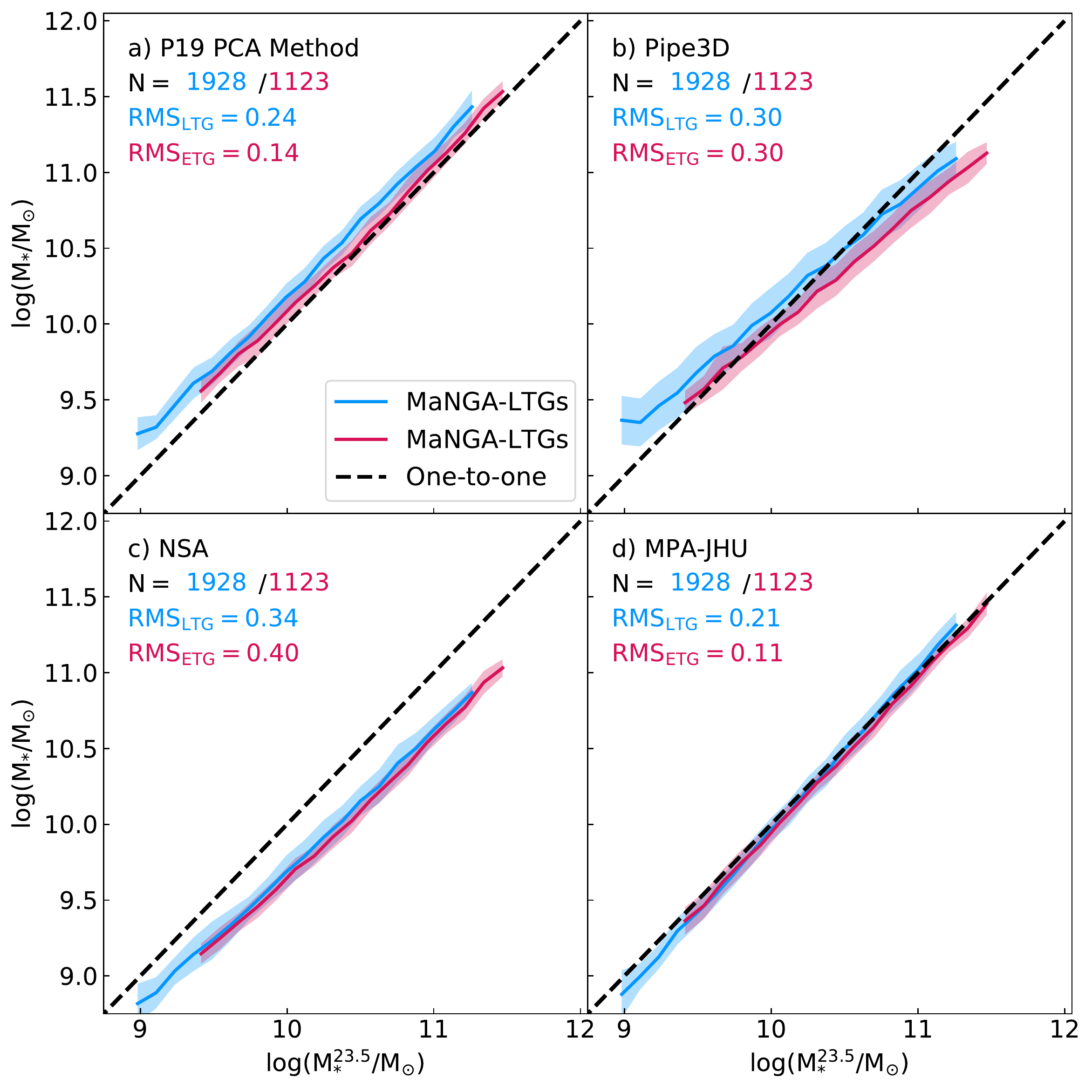}
    \caption{Comparisons of our stellar mass estimates from DESI photometry measured within $\rm R_{23.5,z}$ shown on the {\it x}-axis versus those from various literature sources. The {\it y}-axes compare stellar mass estimates from panel (a): \protect\citetalias{Pace2019}; panel (b): Pipe3D \protect\citep{Sanchez2016}; panel (c): NSA photometry \citep{Blanton2007}; and panel (d) the MPA-JHU catalogue \citep{Kauffmann2003, Brinchmann2004}. The colour scheme matches that of \Fig{f18_size_ser}. The solid line shows the median trends and the shaded regions represent the scatter within a bin.
    The top left inset gives the number of LTG (blue) and ETG (red) galaxies, as well as the rms for the respective stellar mass differences.}
    \label{fig:mstar_compare_b19}
\end{figure*}

\fig{mstar_compare_b19} presents a comparison of our stellar mass estimates on the {\it x}-axis versus various literature sources on the {\it y}-axis.
Our stellar masses are an amalgam of 30 variants derived from optical MLCRs (\citetalias{RC15}; \citetalias{Z17}; \citetalias{Benito2019}).
These MLCRs use stellar population synthesis (SPS) models from \citet[][hereafter BC03]{BC03} and \citet{FSPS}.
The MLCRs presented in \citetalias{Benito2019} apply to all galaxy morphologies, while \citetalias{RC15} and \citetalias{Z17} were only calibrated for LTGs and are here only used for those systems.
The stellar masses from \citetalias{Pace2019} in panel (a) were derived using $\Upsilon_*$ ratios calculated with a principal component analysis (PCA) that finds the best-fitting synthetic spectra for each MaNGA spaxel observation.
Our stellar masses are measured within $R_{\rm 23.5}$ in the {\it z} band, while those from \citetalias{Pace2019} are calculated using the footprint of the MaNGA IFU with aperture correction performed using NSA excess fluxes.

The aperture-corrected stellar masses from \citetalias{Pace2019} appear to be systematically larger than our estimates, especially for MaNGA LTGs (panel a of \fig{mstar_compare_b19}).
The comparison yields an rms offset of 0.25 (0.14) dex and a scatter of 0.09 (0.08) for the LTG (ETG) populations. 
These systematic differences are likely due to the adopted star formation histories (SFHs).
\citetalias{Pace2019} adopted smooth SFH templates; the omission of bursts in SFHs can bias the $\Upsilon_*$ high (\citetalias{RC15}), resulting in larger stellar mass estimates.
These systematic effects are more pronounced for LTGs as these systems have more bursty and active star formation histories. 
However, the reported rms offsets for this comparison are well within the systematic errors expected for MLCRs.
Panel (b) compares our stellar masses against those calculated from Pipe3D \citep{Sanchez2016}; these are integrated within the FOV of the MaNGA IFU.
We find an even larger rms offset and scatter range than our comparison with \citetalias{Pace2019}. 
The Pipe3D comparison shows a large rms difference of 0.30 dex and a scatter of 0.16 (0.1) dex for the LTGs (ETGs).
For both LTGs/ETGs, our stellar mass estimates at the low (high) mass end are smaller (larger) than Pipe3D. 
The difference between our respective stellar mass estimates may arise from $\Upsilon_*$ calculation, which is done pixel by pixel for Pipe3D whereas our $\Upsilon_*$ is calculated using a global colour.

Panel (c) in \fig{mstar_compare_b19} compares stellar mass estimates inferred via {\it K}-corrected elliptical Petrosian photometry from the NSA \citep{Blanton2007}.
In our stellar mass comparisons, NSA estimates present the largest discrepancy with the our photometry. 
Sources for this offset include (i) missing flux in the NSA photometry, (ii) adoption of simple stellar populations (SSPs) for the $\Upsilon_*$
conversions, and/or (iii) the use of Petrosian magnitudes. 
The NSA elliptical-Petrosian photometry has been shown to yield bluer colors $(\Delta (g-r)=0.046\pm 0.008)$ that the SDSS photometry for the complete MaNGA sample \citetalias{Pace2019}, thus causing systematic differences in stellar mass estimates.

Finally, Panel (d) compares our stellar mass estimates with those from the MPA-JHU catalogue \citep{Brinchmann2004}.
Their stellar mass estimates are calculated using the $z$-band stellar $\Upsilon_*$ ratio obtained using the SDSS spectra that best model the H$\,\delta$ and D4000 absorption features applied to $z$ band luminosities.
The comparison in Panel (d) shows a superb match between the MPA-JHU and our photometry with a median across all bins lying close to the 1:1 line and an rms of 0.23 (0.12) dex for the LTGs (ETGs).
Comparing all stellar mass estimates presented in \fig{mstar_compare_b19}, MPA-JHU has the best agreement for both morphological types.

\section{MIR Structural Parameters}
\label{sec:ir_photo}

\begin{figure}
    \centering
    \includegraphics[width=\columnwidth]{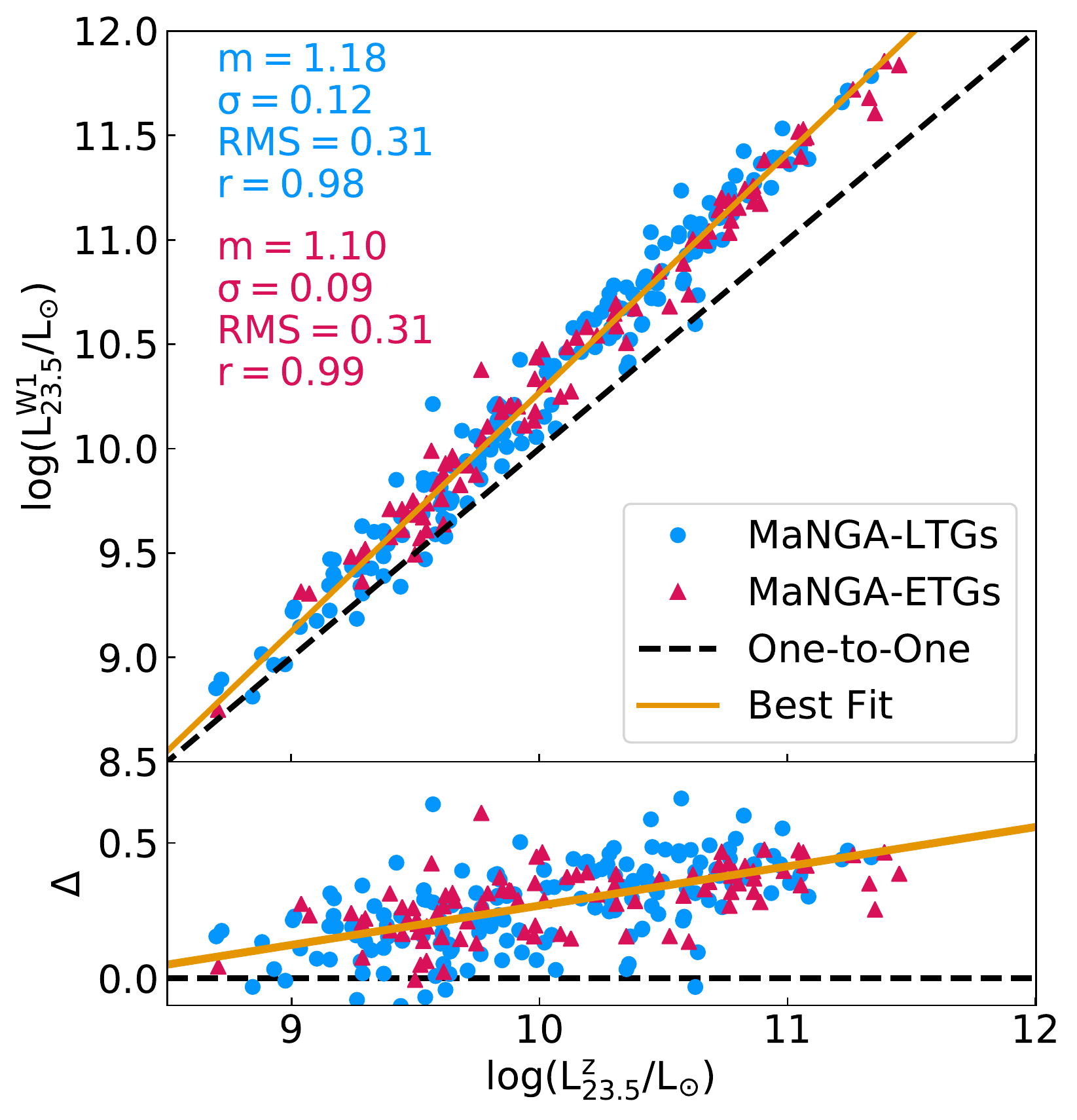}
    \caption{Comparison of total integrated luminosity inferred from DESI and {\it W1} images measured at $R_{23.5,z}$. The blue circles present the MaNGA-LTGs, and red triangles present MaNGA-ETGs. The orange line shows the best fit for the complete population and inset text presents statistics for the two populations.}
    \label{fig:wise_lum}
\end{figure}

\begin{figure}
    \centering
    \includegraphics[width=\linewidth]{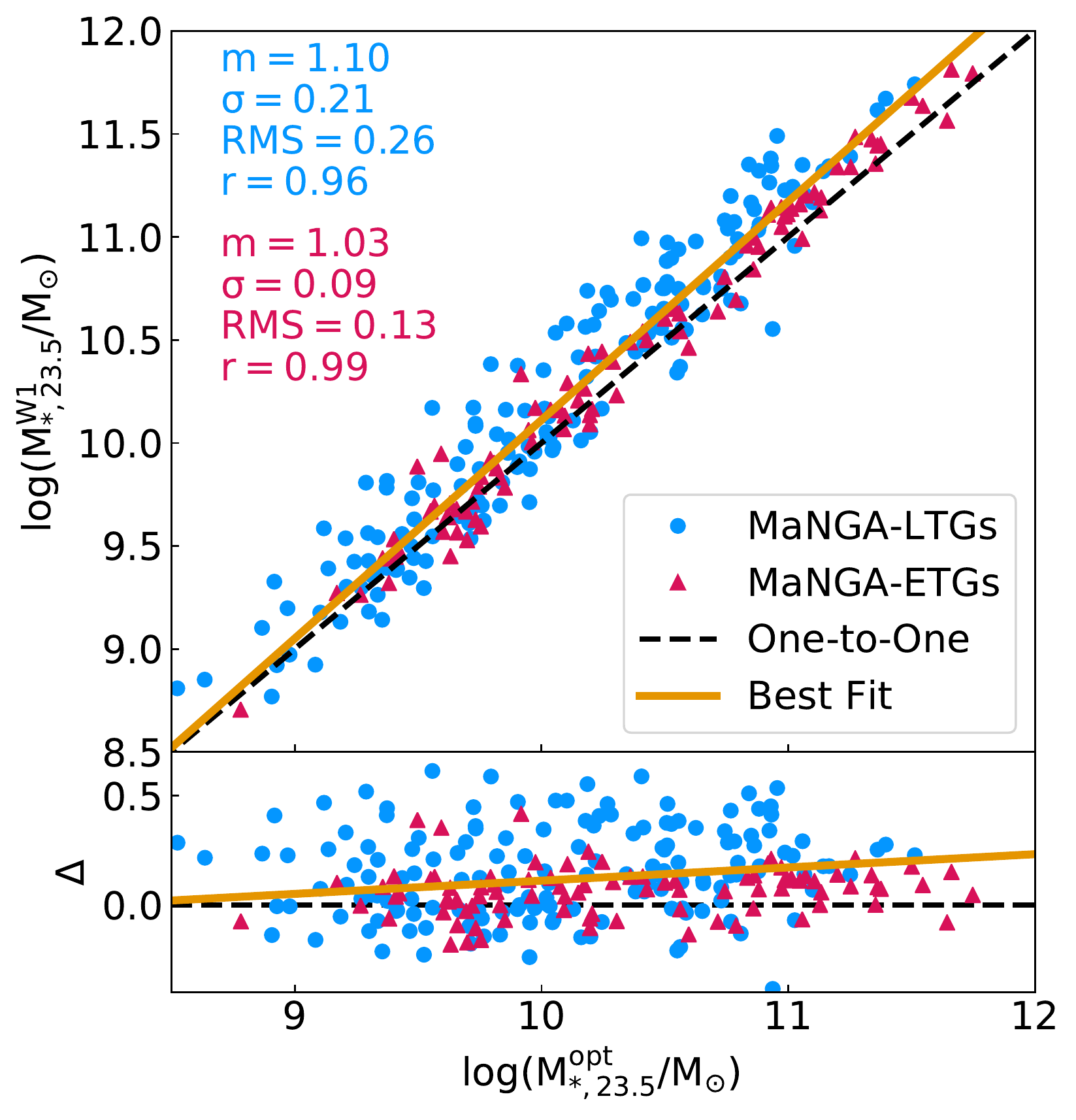}
    \caption{Comparison of stellar masses inferred from DESI and {\it W1} images measured at $R_{23.5,z}$.
    The format is the same as \fig{wise_lum}. For LTGs, "opt" stellar masses are an average of five MLCRs presented in \sec{stellarmass}. For ETGs, "opt" stellar masses are measured using the MLCR of \citetalias{Benito2019}.}
    \label{fig:wise_mstar}
\end{figure}

This section presents a comparison of galaxy structural parameters extracted from our optical and MIR photometry.
We first caution that surface brightnesses sampled with different pixel resolutions, as is the case here with the DESI and WISE data, cannot be compared directly unless profiles are all sampled (degraded) to the lowest resolution. Therefore, only integrated quantities between DESI and WISE data are compared below.

\Fig{wise_lum} shows the {\it W1} and DESI {\it z}-band luminosities measured at $R_{23.5,z}$.
For the complete sample, {\it W1} luminosities are typically larger than {\it z}-band luminosities by $\sim$0.3 dex, as a result of a greater sensitivity of the {\it W1} bandpass to the dominant low mass (older) stellar population and lesser dust extinction in the {\it W1} band.
Furthermore, the {\it W1}-$z$ colour term grows significantly with luminosity, as seen in the residual panel of \Fig{wise_lum}.
The scatter between DESI {\it z} and {\it W1} luminosities is tighter for the MaNGA-ETGs (than LTGs) largely due to their star formation activity being quenched \citep{Cluver2014}. 
The larger scatter for the LTGs population is also indicative of a more diverse stellar population and larger dust content. 

Finally, \Fig{wise_mstar} compares stellar masses inferred from optical and MIR photometry measured at $R_{23.5,z}$. 
As discussed in \sec{stellarmass}, we used the average of five different MLCRs with $g-r$ and $g-z$ colours to calculate optical stellar masses. 
Mass estimates of a stellar population should not depend on the flux tracer, however dust extinction in optical bands could lead to systematic offsets, especially in late-type systems. 
While a systematic offset of 0.21 dex is detected in \fig{wise_mstar}, with {\it W1} stellar masses being larger than optical, the latter is within the bounds of typical MLCRs systematic variations of 0.2-0.3\,dex \citep{Taylor2011,Courteau2014,RC15}. 
Our optical and MIR stellar mass measurements are thus in good agreement~\citetalias{RC15}. 
This reaffirms the conclusion that optical colours are robust tracers of the stellar mass \citep{Taylor2011}.

The differences in luminosity, and stellar mass (\ref{fig:wise_lum} and \ref{fig:wise_mstar} respectively) at different wavelengths can serve as calibrators for stellar population models and numerical models of galaxy formation \citep{MacArthur2004, Renzini2006, Zheng2020}. 

\section{Galaxy Scaling Relations} \label{sec:sr}

In the previous sections, we have demonstrated the robustness of our optical and MIR photometry.
We have also shown that our DESI optical photometry is $\sim$1.5 \magss deeper than nominal SDSS imaging, and we have used the MLDL-VAC to define our MaNGA LTG and ETG subsamples. 

We now present, in \Table{mangaSR}, a variety of scaling relations in DESI ({\it g, r, z}) and WISE ({\it {\it W1}, W2}) band passes for our MaNGA data.
The scaling relations are all presented as orthogonal linear regressions~\citep[][{\scriptsize SCIPY}]{2020SciPy}. 
This treatment facilitates comparisons with matching studies in the literature, though we note that subtle differences in choice of fitting method and parameter definition can affect the final scaling relation parameters (see \Table{SRcompare}). 
As discussed in \sec{param_dist} and \fig{manga_bimodal} some distributions may call for more complex modelling, such as that characterized by a slowly varying slope or a piece-wise function. 
We revisit these complexities below. 

Below we focus on two popular scaling relations drawn from our DESI and WISE photometric investigations of MaNGA LTGs and ETGs, namely the size--stellar mass ($R-M_*$) and $\Sigma_{1}$--stellar mass ($\Sigma_{1}$-M$_*$) relations. 
In the following sections, all structural parameters are measured at R$_{23.5}$ in the $z$ band unless otherwise stated. 

\begin{table*}
\begin{center}
\begin{tabular}{@{}ccccccccc@{}}
\toprule
Scaling relation                & Band      & $y$                       & $x$                   & Type  & Slope         & Zero-point       & Scatter        & N    \\ 
(1)                             & (2)       & (3)                       & (4)                   & (5)     &(6)            & (7)              & (8)            & (9)  \\ \midrule
Projected size--luminosity       &  {\it g}  & $\log R_{23.5}^{g}$       & $\log L_{23.5}^{g}$   & LTGs    & $0.44\pm0.01$ & $-3.55\pm0.03$   & $0.08\pm0.01$  & 2411 \\
                                &           &                           &                       & ETGs    & $0.49\pm0.01$ & $-4.05\pm0.04$   & $0.06\pm0.01$  & 1796 \\
                                &           &                           &                       & All     & $0.42\pm0.01$ & $-3.32\pm0.03$   & $0.10\pm0.01$  & 4215 \\ \hline
Projected size--luminosity       &  {\it r}  & $\log R_{23.5}^{r}$       & $\log L_{23.5}^{r}$   & LTGs    & $0.43\pm0.01$ & $-3.40\pm0.03$   & $0.08\pm0.01$  & 2426 \\
                                &           &                           &                       & ETGs    & $0.52\pm0.01$ & $-4.37\pm0.04$   & $0.06\pm0.01$  & 1801 \\
                                &           &                           &                       & All     & $0.44\pm0.01$ & $-3.40\pm0.03$   & $0.10\pm0.01$  & 4236 \\ \hline
Projected size--luminosity       &  {\it z}  & $\log R_{23.5}^{z}$       & $\log L_{23.5}^{z}$   & LTGs    & $0.41\pm0.01$ & $-3.21\pm0.03$   & $0.08\pm0.01$  & 2456 \\
                                &           &                           &                       & ETGs    & $0.54\pm0.01$ & $-4.56\pm0.05$   & $0.07\pm0.01$  & 1790 \\
                                &           &                           &                       & All     & $0.44\pm0.01$ & $-3.41\pm0.03$   & $0.10\pm0.01$  & 4254 \\ \hline
Projected size--luminosity       &  {\it W1} & $\log R_{23.5}^{{\it W1}}$      & $\log L_{23.5}^{{\it W1}}$  & LTGs    & $0.36\pm0.01$ & $-2.49\pm0.12$   & $0.10\pm0.01$  & 168 \\
                                &           &                           &                       & ETGs    & $0.46\pm0.02$ & $-3.59\pm0.17$   & $0.09\pm0.01$  & 97 \\
                                &           &                           &                       & All     & $0.40\pm0.01$ & $-2.98\pm0.11$   & $0.11\pm0.01$  & 265 \\ \hline
Projected size--luminosity       &  {\it W2} & $\log R_{23.5}^{W2}$      & $\log L_{23.5}^{W2}$  & LTGs    & $0.34\pm0.02$ & $-2.32\pm0.22$   & $0.10\pm0.02$  & 67 \\
                                &           &                           &                       & ETGs    & $0.42\pm0.03$ & $-3.19\pm0.32$   & $0.08\pm0.02$  & 32 \\
                                &           &                           &                       & All     & $0.37\pm0.02$ & $-2.63\pm0.17$   & $0.10\pm0.01$  & 99 \\ \hline
Projected size--stellar mass     &  {\it g}  & $\log R_{23.5}^{g}$       & $\log M_{*, 23.5}$    & LTGs    & $0.34\pm0.01$ & $-2.61\pm0.03$   & $0.12\pm0.01$  & 2408 \\
                                &           &                           &                       & ETGs    & $0.40\pm0.01$ & $-3.39\pm0.04$   & $0.07\pm0.01$  & 1790 \\
                                &           &                           &                       & All     & $0.30\pm0.01$ & $-2.22\pm0.03$   & $0.14\pm0.01$  & 4206 \\ \hline
Projected size--stellar mass     &  {\it r}  & $\log R_{23.5}^{r}$       & $\log M_{*, 23.5}$    & LTGs    & $0.34\pm0.01$ & $-2.57\pm0.03$   & $0.10\pm0.01$  & 2408 \\
                                &           &                           &                       & ETGs    & $0.46\pm0.01$ & $-3.88\pm0.04$   & $0.07\pm0.01$  & 1790 \\
                                &           &                           &                       & All     & $0.34\pm0.01$ & $-2.53\pm0.03$   & $0.12\pm0.01$  & 4206 \\ \hline
Projected size--stellar mass     &  {\it z}  & $\log R_{23.5}^{z}$       & $\log M_{*, 23.5}$    & LTGs    & $0.35\pm0.01$ & $-2.57\pm0.03$   & $0.10\pm0.01$  & 2408 \\
                                &           &                           &                       & ETGs    & $0.49\pm0.01$ & $-4.17\pm0.05$   & $0.07\pm0.01$  & 1790 \\
                                &           &                           &                       & All     & $0.37\pm0.01$ & $-2.73\pm0.03$   & $0.12\pm0.01$  & 4206 \\ \hline
Projected size--stellar mass     & {\it W1}  & $\log R_{23.5}$           & $\log M_{*, 23.5}$    & LTGs    & $0.34\pm0.01$ & $-2.51\pm0.13$   & $0.11\pm0.01$  & 168  \\
                                &           &                           &                       & ETGs    & $0.46\pm0.02$ & $-3.59\pm0.16$   & $0.10\pm0.01$  & 97   \\
                                &           &                           &                       & All     & $0.41\pm0.01$ & $-2.98\pm0.11$   & $0.12\pm0.01$  & 265  \\ \hline
Projected size--stellar mass     & {\it W2}  & $\log R_{23.5}$           & $\log M_{*, 23.5}$    & LTGs    & $0.40\pm0.02$ & $-2.80\pm0.21$   & $0.10\pm0.01$  & 67  \\
                                &           &                           &                       & ETGs    & $0.44\pm0.02$ & $-3.38\pm0.35$   & $0.10\pm0.02$  & 32   \\
                                &           &                           &                       & All     & $0.42\pm0.02$ & $-3.04\pm0.19$   & $0.11\pm0.01$  & 99  \\ \hline
Physical size--stellar mass      & {\it z}   & $\log r_{23.5}$           & $\log M_{*, 23.5}$    & LTGs    & $0.36\pm0.01$ & $-2.83\pm0.04$   & $0.11\pm0.01$  & 2433 \\
                                &           &                           &                       & ETGs    & $0.50\pm0.01$ & $-4.07\pm0.03$   & $0.07\pm0.01$  & 1839 \\
                                &           &                           &                       & All     & $0.42\pm0.01$ & $-3.26\pm0.03$   & $0.10\pm0.01$  & 4274 \\ \hline
Physical size--stellar mass      & {\it W1}  & $\log r_{23.5}$           & $\log M_{*, 23.5}$    & LTGs    & $0.36\pm0.01$ & $-2.51\pm0.12$   & $0.11\pm0.01$  & 168  \\
                                &           &                           &                       & ETGs    & $0.46\pm0.02$ & $-3.45\pm0.16$   & $0.10\pm0.01$  & 97   \\
                                &           &                           &                       & All     & $0.41\pm0.01$ & $-3.02\pm0.12$   & $0.12\pm0.01$  & 265  \\ \hline
$\Sigma_{1}$--stellar mass       & {\it grz} & $\log \Sigma_{1}$         & $\log M_{*, 23.5}$    & LTGs    & $0.96\pm0.01$ & $-0.86\pm0.08$   & $0.24\pm0.01$  & 2408 \\
                                &           &                           &                       & ETGs    & $0.66\pm0.01$ & $\phantom{-}2.43\pm0.11$ & $0.26\pm0.01$  & 1790 \\
                                &           &                           &                       & All     & $0.91\pm0.01$ & $-0.31\pm0.07$   & $0.29\pm0.01$  & 4206 \\ \hline
$\Sigma_{\rm eff}$-stellar mass     & {\it grz} & $\log \Sigma_{\rm eff}$       & $\log M_{*, 23.5}$    & LTGs    & $0.70\pm0.01$ & $\phantom{-}1.16\pm0.13$ & $0.36\pm0.01$  & 2408\\
                                &           &                           &                       & ETGs    & $-0.01\pm0.03$ & $\phantom{-}8.85\pm0.29$& $0.33\pm0.01$  & 1790 \\
                                &           &                           &                       & All     & $0.62\pm0.01$ & $\phantom{-}2.06\pm0.12$& $0.43\pm0.01$  & 4206 \\ \bottomrule
\end{tabular}
\end{center}
\caption{MaNGA Photometric Galaxy Scaling Relations in the $g,r,z,$ {\it W1}, and W2 bands, with orthogonal linear fit results. Columns (1) and (2) give the scaling relation and the relevant photometric band(s); columns (3) and (4) give the variables on the y and x axis coordinates for the scaling relation; column (5) gives the morphological type used for the fit; columns (6)--(8) give the slope (m), zero-point (zp), and scatter ($\rm \sigma$) for our linear orthogonal distance regression; and column (9) gives the number of data points used in each fit.}
\label{tab:mangaSR}
\end{table*}

\begin{table*}
\begin{tabular}{@{}ccccccccc@{}}
\toprule
Scaling relation   & Source               & Sample       & Type & Band & N            & Slope                   & Scatter                 & Size metric    \\
(1)                & (2)                  & (3)          & (4)  & (5)  & (6)          & (7)                     & (8)                     & (9) \\\midrule
$R-M_*$            & This Work            & MaNGA        & LTGs & z    & 2408         & $0.35\pm0.01$           & $0.10\pm0.01$           & $R_{\rm 23.5}$ \\
                   & \cite{Shen2003}      & SDSS         & LTGs & z    & 99\,786       & $0.40$                  & $0.15$                  & $R_{\rm eff}$  \\
                   & \cite{Pizagno2005}   & SDSS         & LTGs & i    & 81           & $0.242\pm0.030$         & n/a                     & $R_{\rm d}$      \\
                   & \cite{Lange2015}     & GAMA         & LTGs & z    & 6151         & $0.21\pm0.02$           & n/a                     & $R_{\rm eff}$  \\
                   & \cite{Ouellette2017} & SHIVir       & LTGs & i    & 69           & $0.34\pm0.02$           & $0.15$                  & $R_{\rm 23.5}$ \\
                   & \cite{Stone2020}     & PROBES       & LTGs & z    & 1152         & $0.334^{0.009}_{0.004}$ & $0.099^{0.002}_{0.003}$ & $R_{\rm 23.5}$ \\
                   & \cite{Trujillo2020}  & IAC Stripe82 & LTGs & i    & 464          & $0.318\pm0.014$         & $0.087\pm0.05$          & $R_{\rm 23.5}$ \\ \midrule
$R-M_*$            & This Work            & MaNGA        & ETGs & z    & 1790         & $0.49\pm0.01$           & $0.07\pm0.01$           & $R_{\rm 23.5}$ \\
                   & \cite{Shen2003}      & SDSS         & ETGs & z    & 99\,786       & $0.56$                  & $0.15$                  & $R_{\rm eff}$  \\
                   & \cite{Lange2015}     & SDSS         & ETGs & z    & 2248         & $0.46\pm0.02$           & n/a                     & $R_{\rm eff}$  \\
                   & \cite{Ouellette2017} & SHIVir       & ETGs & i    & 121          & $0.35\pm0.04$           & $0.12$                  & $R_{\rm 23.5}$ \\
                   & \cite{Trujillo2020}  & IAC Stripe82 & ETGs & i    & 279          & $0.453\pm0.011$         & $0.042\pm0.004$         & $R_{\rm 23.5}$ \\ \midrule
$\Sigma_1-M_*$     & This Work            & MaNGA        & LTGs & grz  & 2408         & $0.96\pm0.01$           & $0.24\pm0.01$           & $R_{\rm 23.5}$ \\
                   & \cite{Barro2017}     & CANDELS      & LTGs & --    & 1328         & $0.89\pm0.03$           & $0.25$                  & n/a              \\
                   & \cite{Woo2019}       & MaNGA        & LTGs & i    & $\sim$41\,000  & 0.86                    & 0.24                    & n/a              \\
                   & \cite{Stone2020}     & PROBES       & LTGs & z    & 1152         & $1.004^{0.021}_{0.035}$ & $0.231^{0.005}_{0.006}$ & $R_{\rm 23.5}$ \\ \midrule
$\Sigma_1-M_*$     & This Work            & MaNGA        & ETGs & grz  & 1790         & $0.66\pm0.01$           & $0.25\pm0.01$           & $R_{\rm 23.5}$ \\
                   & \cite{Barro2017}     & CANDELS      & ETGs & --    & 151          & 0.65                    & 0.14                   & n/a              \\
                   & \cite{Woo2019}       & MaNGA        & ETGs & i    & $\sim$15\,000  & 0.75                    & 0.17                    & n/a              \\
                   & \cite{Fang2013}      & GALEX/SDSS   & ETGs & i    & 1247         & 0.64                    & 0.16                    & n/a
                   \\ \bottomrule
\end{tabular}
\caption{Literature comparisons for scaling relation from our photometry. Column (1) shows the scaling relation; column (2) presents the literature source; column (3) shows the sample used for the scaling relation; column(4) presents the morphological type; column (5) lists the photometric band used for the analysis; columns (6), (7) and (8) present the number of data points, slope and scatter for the best fit line; finally column (9) presents the size metric used for constructing the scaling relation. "n/a" are shown where the data are not available.}
\label{tab:SRcompare}
\end{table*}

\subsection{Size--stellar mass \texorpdfstring{($R-M_*$)}{re} relation}
\label{sec:sizemass}

\begin{figure*}
    \centering
    \includegraphics[width=\linewidth]{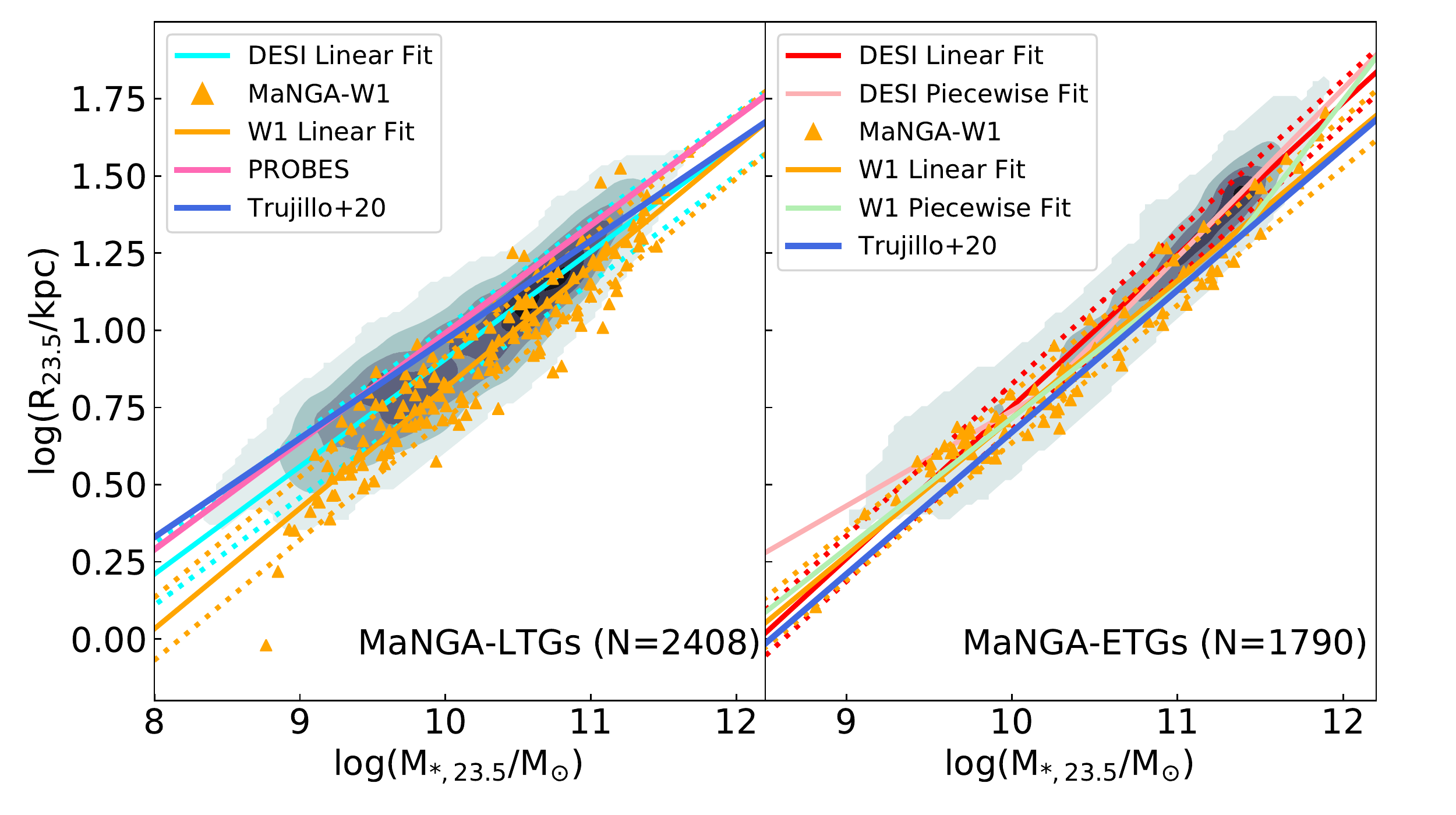}
    \caption{Size--stellar mass relation for LTGs (left-hand panel) and ETGs (right-hand panel). 
    In each panel, grey filled contours represent the distribution of our DESI photometric data; the cyan (LTGs) and red (ETGs) lines are the orthogonal best fit for the corresponding populations. 
    The orange triangles and solid line represent the same structural parameters measured with the WISE ({\it W1}) data. 
    The dotted lines in each panel correspond to the 1$\sigma$ scatter about the best-fitting line. 
    In the left-hand panel, the $R-M_*$ relation for ${\sim}$1100 PROBES galaxies is presented as a pink line \protect\citep{Stone2020}. 
    Both panels show the $R-M_*$ relation of \protect\cite{Trujillo2020} based on ${\sim}$1000 galaxies. 
    Piece-wise linear fits to the DESI and WISE data are also shown for the ETGs (see the text for details.)}
    \label{fig:size_mstar}
\end{figure*}

\Fig{size_mstar} shows the $R-M_*$ relation for LTG and ETG populations measured from our DESI and WISE photometry. 
The parameter distributions are shown as density contours, while the cyan (red) lines represent the orthogonal best fits to the LTG (ETG) data.
The pink solid line shows the best-fitting relation from \cite{Stone2020} who used a compilation of LTG surveys with DESI imaging for $\sim$1100 spiral galaxies.
The photometric analyses for our MaNGA and PROBES samples use the same methods (DESI photometry, \ap, surface brightness profile treatments, photometric band). 
It is therefore not surprising that our linear regression should match the PROBES fit so well. 
\cite{Stone2020} also report a scatter of $0.099^{+0.002}_{-0.003}$ which is in perfect agreement with the observed scatter of our MaNGA galaxies.
Overall, the MaNGA and PROBES LTGs have matching $R-M_*$ properties measured from a distinct sample of galaxies (though matching analysis routines).

The left-hand panel of \fig{size_mstar} (blue solid line) also presents a comparison with the linear $R-M_*$ relation from \cite{Trujillo2020} who reported a slope of $0.318\pm0.014$ and a scatter of $0.087\pm0.005$, based on SDSS-{\it i} band imaging and using isophotal sizes at 23.5\,\magss. 
The different photometric band used in our respective analysis may explain the slight slope and scatter mismatch for our respective $R-M_*$ relations.

For LTGs, the slope of the $R-M_*$ relation is independent of wavelength (\tab{mangaSR}). 
This agrees with \cite{Lange2015} who showed minimal variations in the slope of the $R-M_*$ relation for 8400 GAMA galaxies \citep{gama} across the {\it ugrizZYJHK} photometric bands modulo random variations (see \Table{SRcompare}).
The actual value of our respective slopes differ on account of our respective size metric choices (effective vs. isophotal radius).

The density map and the red solid line in the right-hand panel of \fig{size_mstar} represent the $R-M_*$ relation for our ETG sample.
The single linear $R-M_*$ relation for ETGs is steeper (larger slope) and tighter (smaller scatter) than for LTGs.
The slope and scatter of the ETGs $R-M_*$ relation is likely controlled by their formation via repeated mergers on high impact parameter orbits \citep{Shen2003}.
Dry (gas-less) mergers are the most plausible explanation for the steep size growth of the ETGs From $z \sim 1$ to now \citep{Company2013}.

Our linear $R-M_*$ relation is in general agreement with \citep{Trujillo2020}, who found a slope of $0.453\pm0.011$ and an observed scatter of $0.042\pm0.004$ for 280 galaxies.
The differences in best-fitting parameters are a result of the choice photometric band (SDSS-{\it i}) and least-square linear regression.
\cite{Lange2015} found a range of $R-M_*$ relation slopes of $0.46-0.56$ (scatter not reported) for GAMA galaxies \citep{gama} in the SDSS-{\it z} band. 
They also used effective radii as the size metric making comparisons challenging.
The exact value of the slope in \cite{Lange2015} depends largely on their morphological assignments of ETGs as guided by Se\'rsic index, dust-corrected optical colour, and visual classification.
We find a general trend that the $R-M_*$ relation for ETGs is steeper (shallower) if effective (isophotal) radii are used. 

\fig{size_mstar} also presents the linear $R-M_*$ relation for our WISE {\it W1} photometry with the orange line and triangles for the LTGs/ETGs in the left-/right-hand panels.
This relation (slope and scatter) is essentially the same as that inferred from DESI $z$-band images for our LTGs, further highlighting our conclusions in \tab{mangaSR} that the slope and scatter for LTG $R-M_*$ relation is independent of the wavelength.
The scatter and slope of the ETG WISE $R-M_*$ relation differ slightly from their $z$-band counterpart.
This result is similar to that of \cite{Lange2015}, who found large random variations in the slope of $R-M_*$ relation for ETGs.

Along with the projected $R-M_*$ relation (\tab{mangaSR}), we can calculate this relation with physical radii (see \sec{size}). 
\tab{mangaSR} presents the physical $r_{23.5}-M_*$ relation for the DESI {\it z} and {\it W1} photometric bands.
The projected and physical $r-M_*$ relations for LTGs are equal by definition. 
For ETGs physical sizes increase by 4/3 relative to projected values \citep{Hernquist1990, Chiosi2002}, resulting in a predictable change in zero-points (by 0.125\,dex) of the best-fitting relation.
The best linear fits to physical relations are presented in \tab{mangaSR}. 

\begin{table}
\begin{center}
\begin{tabular}{@{}cccc@{}}
\toprule
Band                          & N    & M$_*$ limit             & Piece-wise fits                   \\ 
(1)                           & (2)  & (3)                     & (4)                               \\ \midrule
DESI-{\it z}                  & 1854 & $\rm \log M_*\leq10.20$ & $\rm 0.31(\log M_* -10.20)+0.80$  \\
                              &      & $\rm \log M_* >10.20$   & $\rm 0.55(\log M_* -10.20)+0.80$  \\ \hline
{\it W1}                      & 97   & $\rm \log M_*\leq10.81$ & $\rm 0.41(\log M_* - 10.81)+1.05$ \\
                              &      & $\rm \log M_*>10.81$    & $\rm 0.52(\log M_* - 10.81)+1.05$ \\ \bottomrule
\end{tabular}
\end{center}
\caption{Linear piece-wise $R-M_*$ relations for MaNGA-ETGs. columns (1) and (2) give the photometric band and number of data points used for the fit; column (3) shows the stellar mass limit for the change in slope; and column (4) presents the piece-wise fits.}
\label{tab:piecewise}
\end{table}

Given the bimodalities in stellar mass and sizes for ETGs seen in \fig{manga_bimodal}, we also model the $R-M_*$ relation based on DESI and WISE imaging with a double piece-wise linear fit (see \tab{piecewise}). It is found that the slopes of our $R-M_*$ relations for LTGs and ETGs based on DESI and WISE data agree closely with the theoretical expectations.
For LTGs, it can be shown that $R_{opt}\propto M_{*}^{0.33}$ for virialized galaxy discs and assuming constant $\Upsilon_*$ per galaxy \citep{Courteau2007}.
Various scaling arguments connecting structure formation to primordial density fluctuations \citep{Blumenthal1984} also suggest that massive ETGs ($\log (M_*/M_{\odot}) \geq 10$) have $R\propto M_*^{0.53}$ \citep{Burstein1997, Chiosi2002, Donofrio2020}.
The slope for lower mass ETGs is predicted to be shallower with $R\propto M_*^{0.33}$; closer to that of LTGs in the same mass range \citep{Chiosi2002, Woo2008}.
The piece-wise linear fit slopes for ETGs (\tab{piecewise}), with a low and high mass transition at $\log (M_*/M_{\odot}) \sim 10.2$ matches theoretical expectations (0.33/0.53) in the same stellar mass bins quite closely \citep{Blumenthal1984, Chiosi2002}.  The same statement holds for the WISE data with stellar mass transition being larger than the optical data.
However, this larger stellar mass transition could be attributed to small statistics for low-mass WISE ETGs.

The agreement between the observed and predicted $R-M_*$ slopes is certainly encouraging. 
However, a complete data-model investigation should also match the zero-points, and scatters of observed scaling relations.  
Our MaNGA data base serves as a stepping stone for such a comparison with theoretical and numerical models of galaxy formation and structure. 

In the next section, we explore other size metrics and their effects on slopes and scatters of the $R-M_*$ relation.

\subsection{Slope and scatter variations of the \texorpdfstring{$R-M_*$}{RM} relation}

\begin{figure*}
    \centering
    \includegraphics[width=\linewidth]{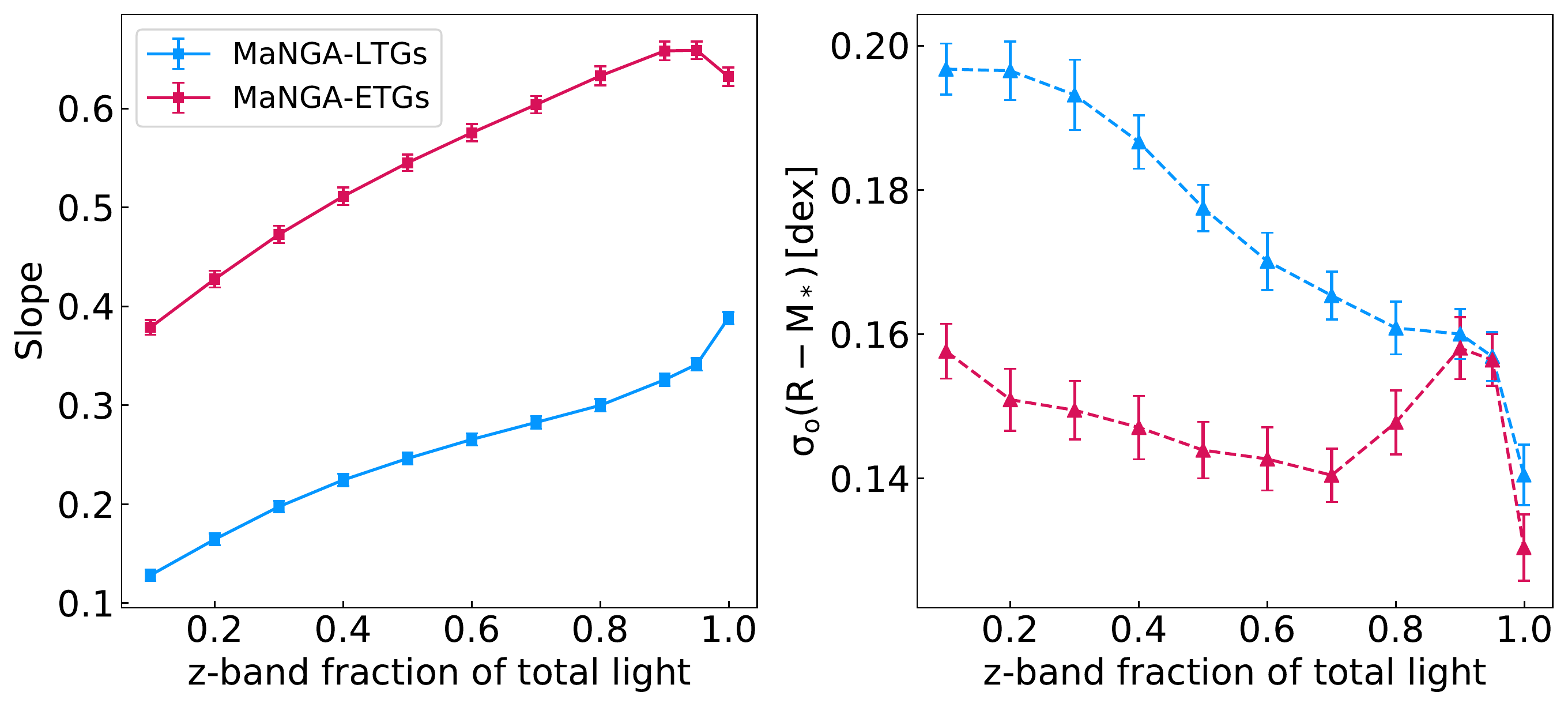}
    \includegraphics[width=\linewidth]{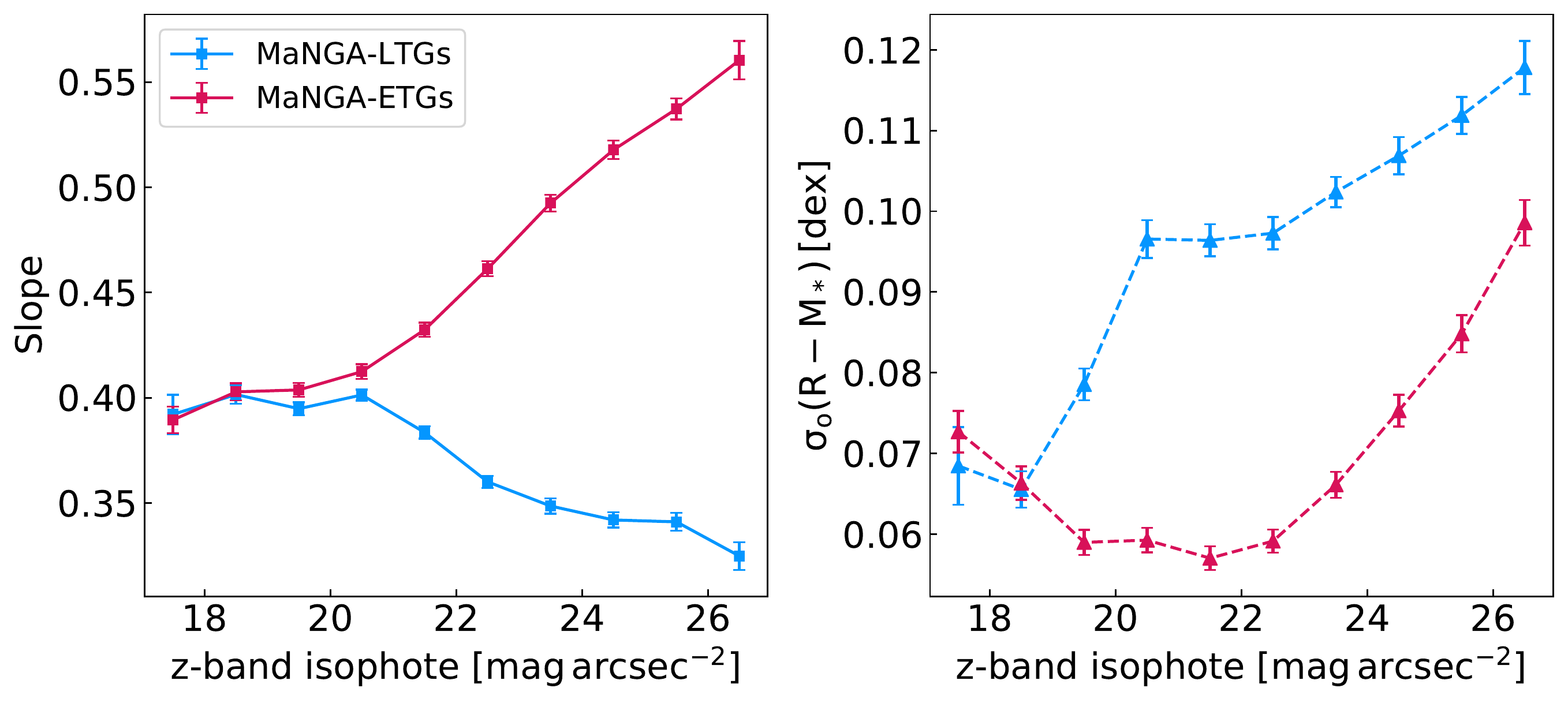}
    \caption{Variations of the slope and orthogonal scatter of the $R-M_*$ relation of LTGs (blue) and ETGs (red) for different size measurements. The top panels show size measurements at various fractions of total light. Isophotal size measurements are presented in the bottom panels. $R-M_*$ slopes are presented in the left-hand panels with connected squares. Observed orthogonal scatter measurements are presented in the right-hand panels with connected triangles. The error bars on all points correspond to bootstrap uncertainties.}
    \label{fig:SizeMassScatter}
\end{figure*}

The spatially-resolved nature of our photometric solutions enables a detailed study of the $R-M_*$ relation with a suite of size metrics for both LTGs and ETGs, as shown in 
\Fig{SizeMassScatter}. 

\subsubsection{Slope variations of the $R-M_*$ relation}

The left-hand panels of \Fig{SizeMassScatter} show variations of the slope of the $R-M_*$ relation with size metrics measured relative to the total light (top panels) and isophotal levels (bottom panels). 
For both morphologies, the slope of the $R-M_*$ relation increases linearly with light fraction.  
However, the $R-M_*$ slopes for LTGs and ETGs with sizes measured at different isophotal levels have nearly opposite behaviours, as seen in the bottom left-hand panel of \fig{SizeMassScatter}.
The $R-M_*$ relation slope for LTGs settles to the theoretical value of 0.33 \citep{Courteau2007} between 23 and 26 $z$-\magss.
The slope for the ETGs grows steadily before flattening to a value of $\sim$0.53 at or beyond 26~$z$-\magss.
The value where the slope flatten agrees with the theoretical prediction found in \cite{Burstein1997} and \cite{Chiosi2002}.
The exact nature of the slope of the $R-M_*$ relation as a function of size metric can be related to the interplay between the relative slopes of the surface brightness profiles and COG.

For LTGs, $R-M_*$ slopes range from 0.13 to 0.39 (fractional) or 0.39 to 0.33 (isophotal); the slopes are always smaller if measured relative to the total light, and the trends are opposite. 
For ETGs, $R-M_*$ slopes range from 0.42 to 0.62 (fractional) or 0.39 to 0.63 (isophotal). 
The steeper $R-M_*$ slope for ETGs (see \sec{sizemass}) is expected from a high occurrence of dry mergers and feedback from stars and supermassive black holes \citep{Shen2003, Company2013}.
The use of $R_{\rm eff}$ also yields a closer match to theoretical predictions of the $R-M_*$ slope. 
The theoretical prediction can also be matched with isophotal radii by binning galaxies in stellar mass \citep[similar to][]{Chiosi2020}.

\subsubsection{Scatter variations of the \texorpdfstring{$R-M_*$}{Rm} relation}

The right-hand panels of \Fig{SizeMassScatter} show variations in the orthogonal scatter of the $R-M_*$ relation with size metrics measured relative to the total light (top panels) and isophotal levels (bottom panels).
For both morphologies, the $R-M_*$ scatter profiles (top right-hand panel of \fig{SizeMassScatter}) show different behaviours.
The scatter for LTGs decreases monotonically from 0.20 to 0.15\,dex with increasing total light fraction. 
The tightest $R-M_*$ relation is found when all the light from our photometry is taken into account.
For ETGs, the scatter of the $R-M_*$ relation is mostly constant around $\sim$0.14\,dex for all size metrics.

Trends with isophotal sizes differ: smallest orthogonal scatter for the $R-M_*$ relation is found at 18.5 (21.6) $z$-\magss for LTG (ETG) populations.
Our conclusions remain the same if forward scatter was used instead of orthogonal.
This result comes as a surprise in light of the current literature that points to the isophotal radius of 23.5~\magss that minimizes the scatter of VRL scaling relations \citep{Courteau1996, Hall2012}.
More recently, \cite{Trujillo2020} found the radius corresponding to a stellar surface density of $\rm 1\,M_{\odot}\,pc^{-2}$ also shows a tight $R-M_*$ relation along with the added benefit of definition motivated by the physics of star formation in galaxies.
This radius corresponds to a fainter isophote ($\sim$26~\magss) which is larger than the isophotal radius of 23.5~\magss.
Our findings contradict this as the radius giving the tightest $R-M_*$ relation is found to be at 18.5 (21.5) z~\magss for LTGs (ETGs).
\cite{Almeida2020} performed a similar exercise and found the isophotal radius of $\rm 24.7\pm0.5$ $r$-\magss minimizes the scatter. 
The difference between our results are related to the definition of scatter of a linear relation (half of inter-quartile range vs. rms) and the choice of the photometric band (DESI-{\it z} vs. SDSS-{\it r}).

Scatter values are also always smaller for isophotal sizes than light fraction sizes.  
This occurs because light fraction sizes encompass multiple surface brightness levels, thus enhancing the mix of stellar populations at any radius and yielding larger $R-M_*$ scatter values \citep[see e.g.][]{Trujillo2020}.
The results in this section are further corroborated with our MIR photometry that show similar trends [minimum scatter on 17.5(20) \magss for ETGs(LTGs)].
These are not shown in \fig{SizeMassScatter} for simplicity. 
It has been noted the the isophotal level of 23.5~$i$-\magss minimizes scatter of the TFR~\citep{Giovanelli1994, Hall2012}. 
In an upcoming publication, we investigate the radius, if any, that minimizes the scatter of the VRL scaling relations simultaneously.

\subsection{\texorpdfstring{$\Sigma_{1}$}{s1}-stellar mass (\texorpdfstring{$\Sigma_1$--M$_*$}{MS}) relation}\label{sec:sigmamass}

\begin{figure*}
    \centering
    \includegraphics[scale=0.65]{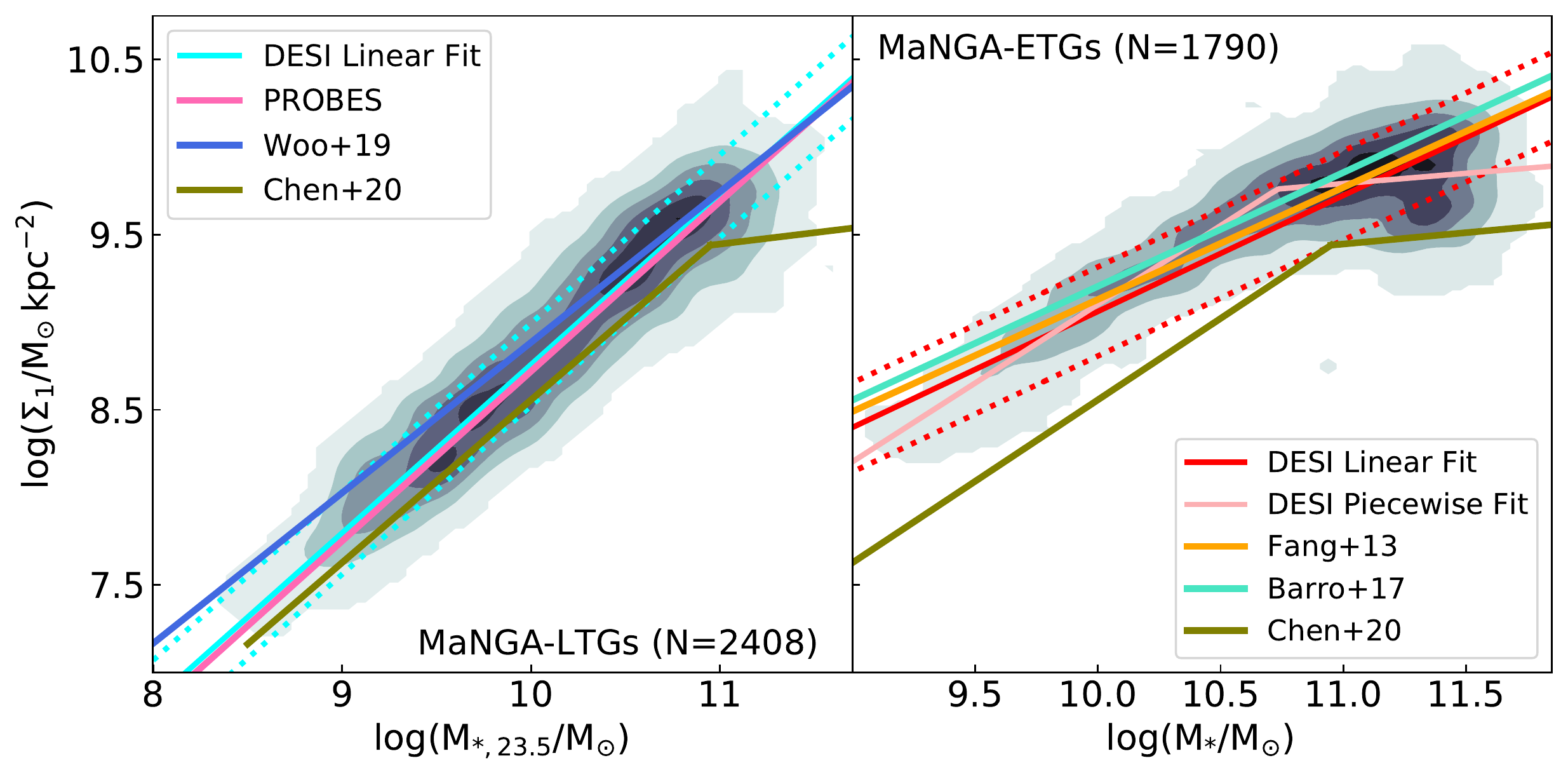}
    \caption{$\Sigma_{1}$ - stellar mass relation for the LTGs (left-hand panel) and ETGs (right-hand panel) using our optical DESI photometry. 
    The density maps in grey show the data distribution. Orthogonal best fits are represented by solid cyan (LTGs) and red (ETGs) lines; the dotted lines show 1\,$\sigma$ scatter for the best-fitting relations. 
    For LTGs, the stellar masses are measured as the average of five MLCRs presented in \sec{stellarmass}. 
    For ETGs, the stellar masses are measured using the MLCR of \protect \citetalias{Benito2019}. 
    We compare our LTG relation with \protect\cite{Woo2019} and \protect\cite{Stone2020}, and our ETG relation with \protect\cite{Fang2013} and \protect\cite{Barro2017} (see the text for the redshift range). 
    Both panels show the \protect$\Sigma_{1}-M_*$ relation of \protect\cite{Chen2020} which fits all morphological types. 
    A piece-wise linear fit to the DESI data is also shown for the ETGs (see the text for details).}
    \label{fig:Mstar_sigma1}
\end{figure*}

The $\Sigma_{1}-M_*$ relation reveals information about the star formation and merger histories of galaxies \citep{Barro2017} and the physical/time evolution of their central components of galaxies \citep{Woo2019, Chen2020, Chen2020b}.
While the calculation of the stellar surface density at 1 kpc can be challenging given large distance errors and saturation issues, $\Sigma_{1}$ traces properties as the bulge component with the added advantage of being model-independent.
\Fig{Mstar_sigma1} shows the $\Sigma_{1} - M_*$ relation for MaNGA LTGs and ETGs based on our DESI optical photometry.
Because the WISE data cannot resolve 1-kpc regions for MaNGA galaxies, this section on $\Sigma_{1}$ only uses our optical photometry.

Our orthogonal fit parameters for the $\Sigma_{1} - M_*$ relation are presented in the bottom row of \tab{mangaSR}.
\cite{Stone2020} found a similar $\Sigma_{1} - M_*$ relation with a slope of $1.005^{0.021}_{0.035}$ and a scatter of $0.23\pm0.01$\,dex for their PROBES sample.

This relation is also fit by \cite{Woo2019} who found a $\Sigma_{1} - M_*$ slope of 0.86 and a scatter of 0.24\,dex using ${\sim}2100$ MaNGA galaxies with SDSS photometry and MPA-JHU stellar masses \citep{Brinchmann2004, Kauffmann2004}. 
Their different slope may be explained by their use of the SDSS {\it i} band, a different definition of stellar mass and a least-square linear fit.  
Their fit is not a good match to our relation (\Fig{Mstar_sigma1}).

Our $\Sigma_{1} - M_*$ slope for ETGs matches closely to that of \cite{Fang2013} whose study of ${\sim}1300$ quenched galaxies selected from the SDSS with $z<0.075$ yielded a slope of $0.64^{+0.23}_{-0.20}$. 
However, their reported scatter of 0.16~dex is significantly smaller than ours.
This disagreement could be due to sample selection, choice of MLCRs to calculate stellar masses and the assumptions about the IMF.
The conversion of light into stellar mass is also inherently uncertain.
While our study targets morphologically selected ETGs, \cite{Fang2013} selected green valley galaxies that are quenched.
Even though there is overlap in these samples, an ETG sample and green valley/quenched sample are different.
For instance, green valley galaxies can exhibit a range of morphologies \citep{Mendez2011} and ETGs can show large range in star formation histories.
In principle, $\Sigma_{1}$ should be more sensitive to the star formation history than visual morphologies explaining our larger scatter for the relation.

The right-hand panel of \fig{Mstar_sigma1} also shows the $\Sigma_{1} - M_*$ relation of \cite{Barro2017} for CANDELS GOODS-S galaxies at $0.5<z<1.0$ \citep{Guo2013} who reported a slope of $0.65\pm0.03$ and a zero-point of $2.71\pm0.05$.
For quiescent galaxies, \cite{Barro2017} found that the slope of the $\Sigma_{1}-M_*$ relation remains constant as a function of redshift; only the zero-point evolves with time.
For a fixed stellar mass bin, $\Sigma_1$ should decrease over time \citep{Barro2017}.
Our similar slopes and smaller zero-point strengthen this assertion as our local universe MaNGA ETGs ($z<0.15$) achieve the same slope and a smaller zero point.

An interesting feature of the $\Sigma_{1} - M_*$ relation for ETGs is its flattening for $\log(M_*/M_{\odot})\geq 10.7$. 
As a result of this feature, and as stated for the $R-M_*$ relation of ETGs \sec{sizemass}, caution should be taken while fitting a linear relation to the ETG $\Sigma_{1}-M_*$ relation (see the right-hand panel of \fig{Mstar_sigma1}).
Along with the linear regression, we fit piece-wise linear function to the data distribution, with $\log\Sigma_1 = 0.90(\log M_* - 10.73)+9.76$ for $\log M_*\leq 10.73$, and $\log\Sigma_1 = 0.12(\log M_* - 10.73)+9.76$ for $\log M_* > 10.73$.
We note that the transition in stellar mass at $\log M_*\sim 10.7$ in this piece-wise fit is mirrored in the bimodal distribution of stellar masses for ETGs seen in \fig{manga_bimodal}.
However, the stellar mass transition in the $R-M_*$ relation of ETGs is found at $\rm \log M_*\sim 10.2$. 
We are reminded that the bivariate distributions are controlled by two random variables.

\cite{Chen2020} also used a piece-wise function to describe the $\Sigma_{1} - M_*$ relation for the complete MaNGA sample calculated to be:
 $\log\Sigma_1 = 0.93(\log M_* - 10.95)+9.44$ for $\log M_*\leq 10.95$, and $\log\Sigma_1 = 0.13(\log M_* - 10.95)+9.44$ for $\log M_* > 10.95$.
Their piece-wise linear fit is shown in the left- and right-hand panels  of \fig{Mstar_sigma1}.
Their slopes for the low- and high-mass ends of the fit match ours quite well.
However, the disagreement in our respective zero-points is significant and may result from our respective stellar mass calculations and different samples.
\cite{Chen2020} used stellar masses provided by the NSA \citep{Blanton2011}; our procedure is described in \sec{stellarmass}. 
Indeed, a large difference is found between the stellar masses by NSA and in this study; our photometry results in larger stellar masses by 0.34 (0.40)\,dex for LTGs (ETGs) that could be due to the systematic offsets of the MLCRs (see \fig{mstar_compare_b19}).
\cite{Chen2020} fit a linear piece-wise function to the full MaNGA sample while our fit is restricted to ETGs.
Both of these effects could cause the observed zero point difference, although it is surprising that the slopes remain unaffected.

For LTGs and low-mass ETGs, the slope of the $\Sigma_{1} - M_*$ relation near unity (0.96 and 0.90) is suggestive of a co-evolution of the inner and outer regions through star formation and environmental interactions.
Indeed, these galaxies may have an enhanced $sSFR(=SFR/M_*)$ which builds up stellar mass in their inner regions \citep{Woo2019}.
The shallower $\Sigma_{1} - M_*$ slope (0.13) at the high-mass end for ETGs likely applies to galaxies with little star formation but ongoing overall accretion, leading to a flattening of the $\Sigma_{1} - M_*$ relation at $\Sigma_1 \sim 10^{10} \rm\,M_{\odot} kpc^{-2}$.
A complimentary explanation for the saturation of $\Sigma_1$ in high mass ETGs involves partially depleted central cores due to coalescing black holes at high redshifts \citep{King1966, Ferrarese1994, Lauer1995, Gebhardt1996, Graham2003}.
A proper appreciation of the saturation of $\Sigma_1$ in galaxies will require additional data, such as SFRs, maps of neutral and molecular gas reservoirs, environmental parameters to characterize interactions and gas infall, and more.
While some of these data still exist, a detailed investigations of the shallower slope at the high-mass end is beyond the scope of this study.
We also caution that $\Sigma_{1}$ may be sensitive to projection, dust extinction, and stellar population effects.

\section{Summary and Future Work} \label{sec:conclusion}

We have presented high-quality optical and MIR surface brightness profiles and environmental properties for the MaNGA galaxy survey.
We made use of DESI imaging and our software (\ap; \citep{Stone2020}) to extract azimuthally averaged optical surface brightness profiles.
On average, the DESI photometry reaches ${\sim}2$\,\magss deeper than the SDSS photometry in the {\it gr} photometric bands which arises from a combination of deeper DESI imaging and our novel technique, \ap.
The WISE profiles are extracted from the WXSC \citep{Jarrett2019} which uses deconvolution techniques to achieve a higher resolution than the native WISE imaging.
70 per cent (33 per cent) of the WISE {\it W1} (W2) surface brightness profiles are as deep in radial extent as the DESI photometry and can be used to compute scaling relations at the fiducial isophotal radius $R_{23.5, z}$.

Excellent agreement is found between most model-independent structural parameters from \ap and those obtained with well-tested surface photometry routines based on the {\scriptsize XVISTA} software package for astronomical image processing \citep{Courteau1996}.
Disagreements between \ap, {\scriptsize XVISTA} and the literature, are largely found for parameters that scale with total light such as effective radii, effective surface brightness, and concentration indices. 
The bimodal nature in the distribution of some structural parameters is also suggestive of distinct galaxy populations in the Universe.

Detailed comparisons of our surface brightness profiles and structural parameters with other studies were presented.
The non-parametric surface brightness profiles from \ap (DESI) and \citetalias{Gilhuly2018} (SDSS) agree well, even reproducing small local variations.
However, the reconstructed surface brightness profiles from the bulge-to-disc decompositions of \citetalias{Gilhuly2018} and \citetalias{Fischer2019} exhibit large differences (${\sim0.4}$\,\magss) demonstrating the challenges involved in such parametric modelling.
Moreover, similar disagreements are found between our model-independent surface brightness profiles and the parametric decompositions of both \citetalias{Gilhuly2018} and \citetalias{Fischer2019}, highlighting once again the fragile nature of parametric modelling.

Our comparisons of effective radii and apparent magnitudes with \pym of \citetalias{Fischer2019} have also revealed disagreements for $R_{\rm eff}$, especially for ETGs $(rms \sim 0.2\,$)dex.
Better comparisons are found for isophotal radii $(rms \sim 0.05\,$)dex, demonstrating the superior reproducibility of isophotal sizes over those measured relative to total light fractions.
Our apparent magnitudes are also typically brighter than those of \citetalias{Fischer2019} by $\sim$0.1\,dex.  
This is expected as, unlike parametric models, our non-parametric surface brightness extraction captures all the light.
The {\scriptsize GALFIT} implementations of \citetalias{Gilhuly2018} and \citetalias{Fischer2019} preferentially favor high S/N regions and systematically predict fainter magnitudes for low S/N.
These systematic effects average out to fainter total integrated light, relative to our non-parametric results.

Our stellar mass estimates, measured at $\rm M_{*}(R_{23.5}^{z})$ and obtained from multiband photometry and various MLCRs, compare favorably with those found in the literature, such as MPA-JHU catalogue \citep{Kauffmann2003}, NSA photometry \citep{Blanton2007}, Pipe3D \citep{Sanchez2016}, and \citetalias{Pace2019}. 
Our stellar masses for LTGs, based on the average of multiple MLCRs from \citetalias{RC15}, \citetalias{Z17}, and \citetalias{Benito2019}, are 0.24\,dex smaller than those of \citetalias{Pace2019}.
This offset is explained by the modelling of SFHs by \citetalias{Pace2019} that systematically biases $\Upsilon_*$ high.
Our stellar masses for ETGs, based on the MLCR from \citetalias{Benito2019}, are 0.11\,dex smaller than those of \citetalias{Pace2019}.
The largest stellar mass differences, found for the NSA stellar mass estimates with an rms of 0.34 (0.40)\,dex for LTGs (ETGs), may stem from uncertainties in the NSA elliptical-Petrosian photometry.
The best match with our stellar mass estimates is found for the MPA-JHU catalogue, with an rms of 0.21 (0.11)\,dex for LTGs (ETGs).

We also present WISE photometry for a subset of $\sim 300$ MaNGA galaxies.
It provides an independent measure of stellar masses, which agree with our optical estimates within 0.21\,dex~\citep[see also][]{Taylor2011}.
Dust extinction may explain the small systematic differences in the stellar mass estimates of LTGs.
In addition to providing accurate stellar masses, spatially-resolved MIR fluxes are most valuable for studies of the star formation main-sequence, SFR$-M_*$ \citep{Cluver2017, Hall2018} stellar population properties \citep{Cluver2014, Cluver2020}.

The slope and scatter of the $R-L_*$ and $R-M_*$ relations for LTGs are found to be independent of bandpass (from {\it g} to {\it W2} \tab{mangaSR}).
For ETGs, slope variations for the $R-M_*$ relation from bluer to redder bands can be linked to varying stellar populations over a range of stellar masses \citep{Lange2015}.
The $R-L_*$ and $R-M_*$ slopes and scatters for LTGs and ETGs also agree well, within 1\,$\sigma$, with published values (see Figures \ref{fig:size_mstar} and \ref{fig:Mstar_sigma1}).

We have examined the variations of the slope and scatter of the $R-M_*$ relation for a range of size metrics.
The slopes of the $R-M_*$ relations for LTGs and ETGs with size metrics measured relative to total light grow linearly with fraction of total light.
Conversely, slopes calculated using isophotal sizes decrease (increase) for LTGs (ETGs) from brighter to  fainter regions.
These trends are dictated by the relative slope variations in the surface brightness profiles and curves of growth.

Isophotal sizes also yield tighter $R-M_*$ relations (smaller scatter) than sizes measured at relative fractions of total light.
The isophotal radius measured at 18.5 (21.5)~\magss also yields the tightest $R-M_*$ relation for the LTG (ETG) population by orthogonal scatter.
Our orthogonal linear fits result in slopes for the LTGs abs ETGs that match theoretical predictions of the $R-M_*$ relations.

The $\Sigma_{1} - M_*$ relation is also presented for both MaNGA LTGs and ETGs and an excellent agreement is found with the literature.
For LTGs, the slope of the $\Sigma_{1} - M_*$ relation is near unity, indicating a co-evolution of stellar mass and $\Sigma_{1}$ that is driven by enhanced star formation and environmental effects.
For ETGs, a near constant $\Sigma_{1}$ is found for $M_{*}>10.5$.
A piecewise linear function was adopted to better match the $\Sigma_{1} - M_*$ distribution. 
The saturation of $\Sigma_{1}$ at high stellar mass could be related to SFRs, environment, neutral/molecular gas distribution, etc.

The multiband photometry, environmental parameters, and structural scaling relations presented in our study may be used to constrain stellar populations models, test semi-analytic or numerical models of galaxy formation, and test their subgrid physics prescriptions \citep{Dutton2009, Brook2012, Henriques2015}.
While extensive, the photometry and scaling relations studied here only inform us about baryonic properties of galaxies.
An essential aspect of galaxy formation and evolution is understanding the co-evolution of baryons and dark matter through a simultaneous study of photometric and kinematic properties of galaxies \citep{Courteau2007, Hall2012, Ouellette2017}.
The MaNGA survey includes dynamical information for all observed galaxies, making this an exciting avenue for related investigations.

\section*{Data Availability}
All data are incorporated into this paper and its online supplementary material.

\section*{Acknowledgements}

We are especially grateful to the Natural Sciences and Engineering Research Council of Canada, Ontario Government, and Queen's University for support through various scholarships and grants.
THJ also acknowledges funding from the National Research Foundation 
under the Research Career Advancement and South African Research 
Chair Initiative programs (SARChI), respectively.
Special thanks go to the referee for valuable suggestions, and to Arjun Dey and the Dark Energy Sky Instrument Legacy Imaging Survey team for their extensive and accessible database.
We also thank Alister Graham for insightful comments that led to further clarifications and judicious references, as well as Stephanie Rendell for contributing to our study of SDSS photometry with {\scriptsize XVISTA}.  Discussions with Mariangela Bernardi enabled a confirmation that the model-dependent \pym ({\scriptsize GALFIT}) apparent magnitudes presented in \citetalias{Fischer2019} are circularized and inclination corrected (assuming infinitesimally-thin transparent discs).
This research made extensive use of astropy~\citep{astropy}, as well as data products from the Wide-field Infrared Survey Explorer, which is a joint project of the University of California, Los Angeles, and the Jet Propulsion Laboratory/California Institute of Technology, funded by the National Aeronautics and Space Administration.



\bibliographystyle{mnras}
\bibliography{reference} 



\appendix

\section{Surface Brightness Profiles}\label{sec:profSB}

This Appendix shows the output format of the DESI-{\it grz} and WISE {\it {\it W1}, W2} surface brightness profiles that are provided as supplementary material.
The DESI and WISE surface brightness profiles are provided in the AB and Vega magnitude systems, respectively.
The conversions from Vega to AB magnitudes in the {\it W1} and {\it W2} bands are given by \citep{Jarrett2013}:  
\begin{equation}
    \rm m_{AB}^{{\it W1}} = m_{Vega}^{{\it W1}} + 2.683,
    \label{eq:w1}
\end{equation}

\begin{equation}
    \rm m_{AB}^{W2} = m_{Vega}^{W2} + 3.319.
    \label{eq:w2}
\end{equation}

\begin{table*}
\begin{tabular}{@{}ccl@{}}
\toprule
Column      & Unit                  & Description  \\
(1)         & (2)                   & (3) \\ \midrule
R           & arcsec                & Semimajor axis length of the isophote  \\
SB          & \magss                & Median surface brightness in the AB magnitude system\\
SB\_E       & \magss                & Error on the surface brightness   \\
MAG         & mag                   & Total magnitude within the isophote, computed by integrating the surface brightness profile \\
MAG\_E      & mag                   & Error on the total magnitude within the isophote \\
ELLIP       & --                   & Ellipticity of the isophote; $\epsilon = 1-b/a$, where $b$ is the semi-minor axis length of the isophote, and $a=R$.   \\
ELLIP\_E    & --                   & Error on the ellipticity, computed by analysing the local variability within 1 PSF \\
PA          & $^{\circ}$                  & Position angle of isophote measured from north to east \\
PA\_E       & $^{\circ}$                   & Error on the position angle, computed by analysing the local variability within 1 PSF \\
MAG\_DIRECT & mag                   & Total magnitude within the isophote computed by flux summation \\
SB\_FIX     & \magss                & Average surface brightness in the AB magnitude system along isophote with ellipticity and position angle set to global values \\
SB\_FIX\_E  & \magss                & Error on SB\_FIX \\
MAG\_FIX    & mag                   & Total magnitude enclosed within the isophote.  Computed by integrating SB\_FIX profile \\
MAG\_FIX\_E & mag                   & Error on MAG\_FIX \\ \bottomrule
\end{tabular}
\caption{Output format of the DESI {\it grz} surface brightness profiles provided as supplementary material. Column (1) refers to the column names in each galaxy surface brightness profile; Column (2) shows the units for each parameter; Column (3) describes each profile entry. All DESI-{\it grz} surface brightness profiles are named with the MaNGA-ID followed with the extension "\_AP.prof"}
\label{tab:desi_profile}
\end{table*}

\begin{table*}
\begin{tabular}{@{}ccl@{}}
\toprule
Column   & Unit     & Description                                                                       \\ 
(1)      & (2)      & (3)                                                                               \\ \midrule
radius   & arcsec   & Semimajor axis length of the isophote                                            \\
SBpix    & --      & Surface brightness in digital number along the isophote                              \\
SBmag    & $\magss$ & Surface brightness in \magss along the isophote in the Vega magnitude system \\
SBerr    & $\magss$ & Error on surface brightness in the Vega magnitude system             \\
SBtotmod & $\magss$ & Total surface brightness for the bulge+disc model in the Vega magnitude system \citep{Jarrett2019} \\
SBbulge  & $\magss$ & Surface brightness of the bulge component in the Vega magnitude system \citep{Jarrett2019}               \\
SBdisk   & $\magss$ & Surface brightness of the disc component in the Vega magnitude system \citep{Jarrett2019}                \\ \bottomrule
\end{tabular}
\caption{Output format of the WISE {\it {\it W1}, W2} surface brightness profiles provided as supplementary material. Column (1) refers to the column names in each galaxy surface brightness profile; Column (2) shows the units for the parameter; Column (3) describes each profile entry. All WISE {\it {\it W1}, W2} surface brightness profiles are named with the WISE catalogue name followed with the extension ".profile.w1/w2.txt". A separate file named "DESI-WISE-index.tbl" provides a conversion table between the MaNGA-ID and WISE-NAME. }
\label{tab:wise_profile}
\end{table*}


\bsp	
\label{lastpage}
\end{document}